\documentclass{article}

\usepackage{algorithm}
\usepackage[margin=30mm,includehead,includefoot]{geometry}
\usepackage[noend]{algpseudocode}
\usepackage{amsmath,amssymb,amsthm,amscd,color,comment}
\usepackage{autobreak}
\usepackage{color}
\usepackage{url}
\usepackage{blkarray}
\usepackage{jheppub_mod}
\usepackage{empheq}
\usepackage{graphicx}
\usepackage{booktabs, multirow}
\usepackage{subcaption}
\usepackage{bm}
\usepackage{enumerate}
\usepackage{microtype}
\usepackage{mathtools}
\usepackage{wrapfig}
\usepackage{caption}
\usepackage{enumitem}

\usepackage{multicol}

\usepackage{tikz}
\usetikzlibrary{patterns}
\usetikzlibrary{arrows,shapes,positioning}
\usetikzlibrary{trees}
\usetikzlibrary{matrix,arrows} 				
\usetikzlibrary{positioning}				
\usetikzlibrary{calc,through}				
\usetikzlibrary{decorations.pathreplacing}  
\usepackage{pgffor}							
\usetikzlibrary{decorations.pathmorphing}	
\usetikzlibrary{decorations.markings}
\tikzstyle{block} = [draw, rectangle, 
minimum height=3em, minimum width=6em]

\tikzstyle{decision} = [diamond, draw, fill=blue!20, 
text width=4.5em, text badly centered, node distance=3cm, inner sep=0pt]
\tikzstyle{block} = [rectangle, draw, fill=blue!10, 
text width=7em, text centered, rounded corners, minimum height=4em, node distance=3cm]
\tikzstyle{line} = [draw, -latex']
\tikzstyle{cloud} = [draw, rectangle, fill=blue!20, text width=6em, text centered, rounded corners, node distance=3cm,
minimum height=2em]
\tikzstyle{border} = [draw, dashed, rectangle, fill=blue!5, rounded corners, node distance=3cm, minimum height=25em, minimum width=22em]
\tikzstyle{data} = [draw, rectangle, fill=blue!10, rounded corners, minimum height=2em, minimum width=8em]
\tikzstyle{box} = [draw, rectangle, fill=white, rounded corners, minimum width=10em]
\tikzstyle{input}=[trapezium, draw, text centered, trapezium left angle=60, trapezium right angle=120, minimum height=2em, fill=blue!10]

\tikzset{fermionnoarrow/.style={draw=black},}

\definecolor{green1}{HTML}{3D792A}
\definecolor{cyan1}{HTML}{37cdaa}
\definecolor{blue1}{HTML}{5d7ac4}
\definecolor{red1}{HTML}{d0482a}
\definecolor{purple1}{HTML}{845ea8}
\definecolor{orange1}{HTML}{e07229}

\title{Gravitoelectric dynamical tides at second post-Newtonian order}

\author[a,b]{Manoj K. Mandal,}
\author[c,a]{Pierpaolo Mastrolia,}
\author[d]{Hector O. Silva,}
\author[d,e]{Raj Patil,}
\author[d]{Jan Steinhoff}

\newcommand{\unipd}{Dipartimento di Fisica e Astronomia, Universit\`a degli Studi di Padova,
Via Marzolo 8, I-35131 Padova, Italy.}

\newcommand{\pdinfn}{INFN, Sezione di Padova,
Via Marzolo 8, I-35131 Padova, Italy.}

\newcommand{\ucla}{Mani L. Bhaumik Institute for Theoretical
Physics, University of California at Los Angeles, Los Angeles, CA
90095, USA.}

\affiliation[a]{\pdinfn}
\affiliation[b]{\ucla}
\affiliation[c]{\unipd}
\affiliation[d]{Max Planck Institute for Gravitational Physics (Albert Einstein Institute), Am M{\"u}hlenberg 1, Potsdam D-14476, Germany}

\affiliation[e]{Institut f{\"u}r Physik und IRIS Adlershof, Humboldt-Universit {\"a}t zu Berlin, Zum Großen Windkanal 2, D-12489 Berlin, Germany}

\emailAdd{manojkumar.mandal@pd.infn.it}
\emailAdd{pierpaolo.mastrolia@unipd.it}
\emailAdd{hector.silva@aei.mpg.de}
\emailAdd{raj.patil@aei.mpg.de}
\emailAdd{jan.steinhoff@aei.mpg.de}

\abstract{ 
We present a gravitoelectric quadrupolar dynamical tidal-interaction Hamiltonian for a compact binary system,  that is valid to second order in the post-Newtonian expansion.
Our derivation uses the diagrammatic effective field theory approach, and involves Feynman integrals up to two loops, evaluated with the dimensional regularization scheme.
We also derive the effective Hamiltonian for adiabatic tides, obtained by taking the appropriate limit of the dynamical effective Hamiltonian, and we check its validity by verifying the complete Poincar\'e algebra. 
In the adiabatic limit, we also calculate two gauge-invariant observables, namely, the binding energy for a circular orbit and the scattering angle in a hyperbolic scattering.
Our results are important for developing accurate gravitational waveform models for neutron-star binaries for 
present and future gravitational-wave observatories.
}

\newcommand{\dd}{{\rm d}}
\newcommand{\DD}{{\rm D}}
\newcommand{\ii}{{\rm i}}

\allowdisplaybreaks

\begin{document}
\addtocontents{toc}{\protect\setcounter{tocdepth}{2}}

\begin{flushright}
\begingroup\footnotesize\ttfamily
	HU-EP-23/09-RTG
\endgroup
\end{flushright}

\maketitle

\section{Introduction}

The detection of gravitational waves (GWs) produced by the inspiral of neutron-star (NS) binaries~\cite{LIGOScientific:2017vwq,LIGOScientific:2017ync,LIGOScientific:2020aai} is a unique probe into the physics of dense nuclear matter inside these stars.
The phasing of the GW signal carries information not only about the binary component's masses, but also about their mutual tidal interaction~\cite{Flanagan:2007ix}. A NS under the influence of its companion's tidal field acquires a quadrupole moment and, depending on the binary's orbital frequency, the NS can have its normal modes of oscillation excited. The magnitude of these two effects depends on star's mass and on the EOS (see Fig.~\ref{fig:bns}.)
The energy spent in deforming each star comes at the expense of the binary's binding energy making
the inspiral dynamics unfold faster.
The imprint of tidal interactions in the GW signal was observed in GW170817~\cite{LIGOScientific:2017vwq} and lead to constraints on 
the underlying NS EOS~\cite{LIGOScientific:2018hze,LIGOScientific:2018cki,Chatziioannou:2020pqz,Pradhan:2022rxs}.

NSs feature a number of oscillation modes, and to
understand them we can picture a basic stellar model that consists of the continuity equation (conservation of mass), Euler's equation (equation of motion for the fluid elements), Poisson's equation (that determines the gravitational force from the matter distribution), and the EOS (that describes how pressure and density are related).
With these elements combined, we can describe a star in equilibrium, which we can 
then perturb.
The resulting \emph{normal modes of oscillation} can be classified as follows~\cite{McDermott:1983ApJ...268..837M,McDermott:1988ApJ...325..725M,christensen1998lecture,Kokkotas:1999bd,Andersson:2021qdq}.

The acoustic waves, known as {\it p-modes}, arise when the equilibrium state of the star is homogeneous. The restoring force is due to pressure, hence their name.
For p-modes, the radial component of the fluid perturbations is usually significantly larger than the tangential component, and these modes are thus sensitive to the compressibility of matter. 
The gravity waves, known as {\it g-modes}, arise when the equilibrium state of the fluid is stratified due to gravity. 
The buoyancy acting on fluid elements provides the restoring force. For g-modes, the tangential component of the fluid perturbations is significantly larger than the radial component. 
We can distinguish the p- and g-modes by their evolution in the phase diagram as one approaches from the center to the star's surface~\cite{Osaki1975,christensen1998lecture}.
When the fluid is assumed to be of constant density, the gravity waves travel only on the surface, and thus do not have any nodes in the radial direction.
These waves are called \emph{surface gravity waves} and their frequency depend only on the mean density of star. Therefore, they are approximately insensitive to the EOS~\cite{Andersson:1995ez,Andersson:1997rn}.\\
\begin{minipage}{.49\textwidth}  
\begin{figure}[H]
\includegraphics[width=1\linewidth]{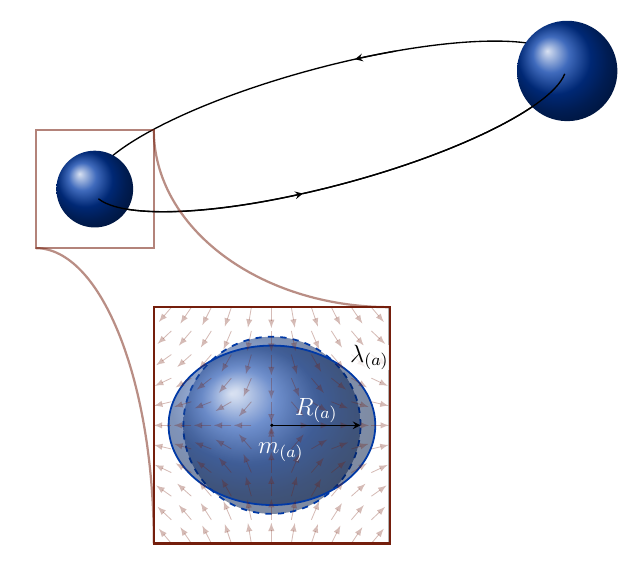}
{\captionsetup{labelformat=empty}
\caption{}}
\end{figure}
\end{minipage}
\begin{minipage}{.5\textwidth} 
\begin{enumerate}[label=\textbf{Figure \arabic*}, ref={\arabic*}, wide, labelwidth=!, labelindent=0pt]
\item Illustration of the problem. Two neutron stars with masses $m_{(a)}$
and radii $R_{(a)}$ ($a = 1, 2$) orbit one another. Each star experiences a tidal 
field due to the gravitational field of its companion.
The tidal field induces a quadrupolar deformation (with magnitude encoded in the tidal Love number $\lambda_{(a)}$) and the displacement away from equilibrium of the star's fluid elements is described as an harmonic oscillator with angular frequency $\omega_{f(a)}$, related to the star's fundamental (f-)mode.
The values of $\lambda_{(a)}$ and $\omega_{f(a)}$ depend on the star's mass and internal composition. The conservative dynamics of this \emph{dynamical tidal problem} is 
studied here to second post-Newtonian order using an effective field theory description.
\label{fig:bns}
\end{enumerate}
\end{minipage}

The lowest frequency surface gravity waves is known as the {\it f-mode}, which is one of the dominant modes in the context of tidal excitation~\cite{Will:1983vlw}.
The relation between orbital motion and the quadrupolar f-modes was first studied by Cowling~\cite{cowlingfmodes} in Newtonian gravity and then in Refs.~\cite{shibatafmodes,Kokkotas:1995xe,Lai:1993di,Ho:1998hq,Lai:2006pr} in general relativity. 
The quadrupole f-mode oscillation of a NS coupled to the external tidal field can be described by the Newtonian Lagrangian (see Ref.~\cite{Flanagan:2007ix} or Ref.~\cite{poisson2014gravity}, Sec.~2.5)
\begin{align}\label{eq_dytide_NRlimit}
{\cal L}_{N}=\frac{1}{4\lambda\omega_f^2} \left[ \frac{\dd\bm{Q}^{ij}}{\dd t}\frac{\dd\bm{Q}^{ij}}{\dd t} -\omega_f^2 \bm{Q}^{ij}\bm{Q}^{ij} \right] -\frac{1}{2} \bm{E}^{ij}\bm{Q}^{ij} \,,
\end{align}
where $\omega_f$ is the frequency of the f-mode and $\lambda$ is the tidal deformability.\footnote{The tidal deformability is related to the dimensionless electric-type quadrupolar
Love number $k_2$ of the body and the radius $R$ of the star as $k_2 = 3 G_N \lambda / (2R^5)$~\cite{Flanagan:2007ix}.
}
While this Lagrangian only describes the f-mode, it phenomenologically provides a very good approximation for the total linear gravitoelectic tidal response, since tidal contributions from other modes (e.g. p-modes) are typically much smaller.
Then $\bm{Q}^{ij}$ is the quadrupole moment of the star and $\bm{E}^{ij}=\partial_i\partial_j\Phi_{\rm ext}$ is the quadrupolar tidal field given
in terms of spatial derivatives of the external Newtonian gravitational potential $\Phi_{\rm ext}$.
In the limit in which $\omega_f \rightarrow \infty$, the Lagrangian Eq.~\eqref{eq_dytide_NRlimit} describes adiabatic tides. In this limit, the tidal bulges do not oscillate, and are instead locked to the external tidal field as $\bm{Q}^{ij} = - \lambda \bm{E}^{ij}$~\cite{Hinderer:2007mb,Binnington:2009bb,Damour:2009vw}.
Qualitatively, the tidal deformability, encoded by the Love numbers, describe how easily a body is deformed in response to external tidal forces~\cite{Love:1911}. The value of the Love numbers depend on the body's internal composition, and as the compactness of the body increases, the value of the Love numbers decrease and eventually approaches zero for a black hole~\cite{Binnington:2009bb} 
(see also Ref.~\cite{Yagi:2016ejg}.)

The relativistic version of the~\eqref{eq_dytide_NRlimit} can be obtained by demanding that the Lagrangian is invariant under Lorentz transformations and reparametrization of worldlines, as first proposed in Ref.~\cite{Steinhoff:2016rfi},
\begin{align}\label{eq_Lag_dytides_1}
\mathcal{L}_{\rm DT}=\frac{z }{4\lambda\omega_f^2} \left[ \frac{c^2}{z^2}\frac{\dd Q_{\mu\nu}}{\dd\tau}\frac{\dd Q^{\mu\nu}}{\dd\tau} -\omega_f^2 Q_{\mu\nu}Q^{\mu\nu} \right] -\frac{z}{2} E_{\mu\nu}Q^{\mu\nu} \,,
\end{align}
where $Q_{\mu\nu}$ is a symmetric trace-free tensor that models the relativistic quadrupole moment of the star, $E_{\mu\nu}= - c^2 R_{\mu\alpha\nu\beta}u^\alpha u^\beta/z^2$ is the gravitoelectric field\footnote{We differ from the action in Ref.~\cite{Steinhoff:2016rfi} for the signature of the metric. Therefore, we add a prefactor of $-c^2$ in the definition of $E_{\mu\nu}$ so that the leading order contribution matches the required result $\bm{E}_{ij}=\partial_i\partial_j\Phi_{\rm ext}$.}, which is the relativistic analogue of the Newtonian external tidal field, $z=\sqrt{u^2}$ is the redshift factor, and $\tau$ is the proper time, related to the coordinate time $t$ as $\dd \tau=c~\dd t$.
Since $Q_{\mu\nu}$ has 9 degrees of freedom, whereas the physical quadrupole of the NS has only 5, we also have to impose a gauge condition
\begin{align}\label{eq_Q_guage}
Q_{\mu\nu}u^\mu=0 \,.
\end{align}
The most notable effects introduced by the relativistic Lagrangian~\eqref{eq_Lag_dytides_1} are the appearance of redshift and frame-dragging effects~\cite{Steinhoff:2016rfi}.
In the relativistic case the adiabatic limit is also obtained by taking $\omega_f\rightarrow \infty$ limit. This limit gives us the equation of motion for $Q_{\mu\nu}$,
\begin{align}\label{eq_Q_eq_E}
Q_{\mu\nu}=-\lambda E_{\mu\nu} \,,
\end{align}
which substituted back into Eq.~\eqref{eq_Lag_dytides_1} results in the Lagrangian for adiabatic tides
\begin{align}\label{eq_Lag_ad}
\mathcal{L}_{\rm AT}=   \frac{z\lambda}{4}  E_{\mu\nu}E^{\mu\nu} \,.
\end{align}
Similarly, we can also write a Lagrangian for the higher adiabatic multipole moments which were studied in Ref.~\cite{Bini:2012gu}. See also Refs.~\cite{Thorne:1984mz,Zhang:1986cpa,Damour:1990pi,Binnington:2009bb,Damour:2009wj,Damour:2009vw,Henry:2019xhg}.
In general relativity, in addition to the relativistic gravitoelectric tides, we also get a new sector of gravitomagnetic tides \cite{Favata:2005da,Landry:2015cva,Pani:2018inf,Banihashemi:2018xfb,Poisson:2020eki,Poisson:2020mdi,Poisson:2020ify,Gupta:2020lnv} that are coupled to the odd-parity normal modes of the NS, modeled by the current-type multipole moments. For the adiabatic limit of the gravitomagnetic sector, see Ref.~\cite{Bini:2012gu}.
We note that Eq.~\eqref{eq_Q_eq_E} is justified here from the phenomenological observation that the f-mode contributes the dominant gravitoelectic tidal effect.
In a systematic EFT construction of dynamical tides, further couplings should we included, as outlined in Ref.~\cite{Gupta:2020lnv} and applied to gravitomagnetic tides, which we leave for future work.

Why should one care about modelling dynamical tidal effects? 
Recently, Ref.~\cite{Pratten:2021pro} showed that the higher-order tidal effects, specifically the f-mode dynamical tides, are important to the inference of the NS EOS with current GW detectors.
The absence of dynamical tidal effects can lead to substantial biases to the inference of the tidal deformability which, in turn, translate into an inaccurate inference of the EOS.
Moreover, the inclusion of dynamical tides are also known to improve the agreement between GW models and numerical relativity simulations~\cite{Steinhoff:2016rfi,Hinderer:2016eia,Andersson:2019dwg,Schmidt:2019wrl}.
Accurate waveform models are also necessary to fulfill the scientific goals of next generation ground-based GW observatories~\cite{Punturo:2010zza,Reitze:2019iox,Ackley:2020atn}.

With these motivations in mind, we examine here how the dynamic tides affect the dynamics of a compact binary. To do so, we use effective field theories (EFT) techniques~\cite{Goldberger:2004jt} to analyze the binary's inspiral, i.e., when the the binary components are moving at nonrelativistic velocities and the orbital separation is large.
In this regime, we can use a perturbative approach that involves a series expansion in powers of $v/c$, where $v$ is the orbital velocity of the binary and $c$ is the speed of light. 
The virial theorem requires that the kinetic to be $-1/2$ times the potential energies of a bound state system.
Hence, we can perform a \emph{post-Newtonian} analysis which involves an expansion in two perturbative parameters: $v/c$ and $G_N$, where $G_N$ is Newton's constant. 
Terms of order $(v/c)^{n}$ are said to be of $(n/2)$PN order.
The PN analysis of the binary dynamics can be divided into two sectors, namely the conservative sector, where the emitted radiation is neglected and the orbital separation does not decrease, and the radiative sector, where the emitted radiation carries away energy and momentum. 
At higher PN orders, these sectors can mix, as due to tail effects which originate from radiation being scattered by the orbital background curvature interacting back onto the orbital dynamics (see, e.g., Ref.~\cite{Blanchet:2013haa}.)
%
Using the EFT approach, we can determine any observable quantity at any given PN order. By using modern EFT diagrammatic based methods, first proposed in Ref.~\cite{Goldberger:2004jt} and modern integration methods \cite{Kol:2013ega,Foffa:2016rgu}, which we recently applied also to account for spin-dependent effects in Refs.~\cite{Mandal:2022nty,Mandal:2022ufb}, the problem is turned into the determination of scattering amplitudes.
These amplitude can be systematically obtained through the calculation of the corresponding Feynman diagrams. See, e.g., Refs.~\cite{Porto:2016pyg, Levi:2018nxp,Goldberger:2022ebt} for reviews. 

The state-of-the-art of the conservative dynamical gravitoelectric tides is the 1PN effective Hamiltonian computed in Refs.~\cite{Vines:2010ca,Steinhoff:2016rfi}. The effects of spin and tides were analyzed together in Refs.~\cite{Gupta:2020lnv,Steinhoff:2021dsn,Gupta:2023oyy} 
for
gravitomagnetic tides. In the adiabatic limit, the 2PN effective Hamiltonian was computed in Refs.~\cite{Bini:2012gu,Henry:2019xhg} for both gravitoelectric and gravitomagnetic tides. 
Other works in PN theory can be found in Refs.~\cite{Vines:2011ud,Bini:2012gu,Vines:2010ca,Abdelsalhin:2018reg,Banihashemi:2018xfb,Landry:2018bil}.
In the post-Minkowskian (PM) expansion, where the perturbative series is controlled by $G_N$ alone, the adiabatic tidal corrections were studied to 3PM order 
in Refs.~\cite{Jakobsen:2022psy,Kalin:2020lmz}.
See also Refs.~\cite{Kalin:2020mvi,Bini:2020flp,Cheung:2020sdj,Haddad:2020que,Cheung:2020gbf,Bern:2020uwk,Mougiakakos:2022sic}.
Adiabatic tidal effects where also included to effective-one-body waveform models~\cite{Buonanno:1998gg,Buonanno:2000ef} in Refs.~\cite{Bini:2012gu,Damour:2009wj,Baiotti:2011am,Bernuzzi:2012ci,Bini:2014zxa} and in Refs.~\cite{Hinderer:2016eia,Steinhoff:2016rfi,Steinhoff:2021dsn} for the case of dynamical tides.

In this paper, we extend the state-of-the art of the \emph{analytic calculations of dynamical gravitoelectric tides in the conservative sector by working to 2PN order}, and we discuss a few physical applications.
The paper is organized as follows. 
In Section~\ref{sec_setup}, we review
the description of tidally-interacting binaries in the EFT formalism.  
Next, in Section~\ref{sec_Routine}, we present the algorithm used to compute the 2PN dynamic tidal potential.
Our main result, the effective dynamical tidal Hamiltonian~\eqref{eq_ham_dy}, is presented in Section~\ref{sec_results}. 
In Section~\ref{sec_adiabatic}, we consider the adiabatic limit, and derive an effective adiabatic tidal Hamiltonian. 
We scrutinize this result by performing a nontrivial check of the Poincar\'e algebra and, as applications, we compute two gauge-independent observables: (i) the binding energy of a circular binary and (ii) the scattering angle for the hyperbolic encounter of two stars. 
Finally, we present our conclusions and avenues for future work in Section~\ref{sec_Conclusion}. 
This work is supplemented with two ancillary files: 
\texttt{Hamiltonian-DT.m}, containing the analytic expression of the Hamiltonian for the dynamic tides and \texttt{Hamiltonian-AT.m}, containing the analytic expression of the Hamiltonian for the adiabatic tides.

{\it Notation --}~The mostly negative signature for the metric is employed. Bold-face characters are used for three-dimensional variables, and normal-face font, for four-dimensional variables. The subscript ${(a)}$ labels the binary components on all the corresponding variables, like their position $\bm{x}_{(a)}$ and quadrupole moment $\bm{Q}_{(a)}$. An overdot indicates the time derivative, e.g., $\bm{v}_{(a)}=\dot{\bm{x}}_{(a)}$ is the velocity, $\bm{a}_{(a)}=\ddot{\bm{x}}_{(a)}$ the acceleration and $\dot{{\bm Q}}= \dd {\bm Q}/\dd t$. The separation between two objects is denoted by $\bm{r}=\bm{x}_{(1)}-\bm{x}_{(2)}$, with absolute value $r=|\bm{r}|$ and the unit vector along the separation is $\bm{n}=\bm{r}/r$.
\section{An EFT description of dynamical tides}
\label{sec_setup}

In this section, we introduce the EFT description of dynamical tides in a compact binary, following closely the presentation in Ref.~\cite{Steinhoff:2016rfi}. This section will also serve to fix the notation that will be used in the remainder of this paper.

We begin by defining three reference frames. These are: 
(i) the general coordinate frame (denoted by Greek indices), 
(ii) the local Lorentz frame (denoted by small Latin indices), 
(iii) and the rest frame of the compact objects (denoted by capital Latin indices). 
The dynamical quadrupolar variables in the different frames are then given by 
\begin{align}
Q_{(a)}^{\mu\nu}=e^{\mu}_{~a}e^{\nu}_{~b}Q_{(a)}^{ab}\,, \quad \text{and} \quad Q_{(a)}^{ab}=B^{a}_{(a)A}B^{b}_{(a)B}Q_{(a)}^{AB}\,,
\end{align}
%
where $e^{\nu}_{~b}$ is the tetrad that transforms between the general coordinate frame and the local Lorentz frame. The Lorentz transformation, which boosts between the local Lorentz frame and the rest frame of the body is given by
\begin{align}\label{eq_boostopt}
B^{a}_{(a)A}=\eta^{a}_{~A} + 2 \frac{u_{(a)}^a \delta_{A}^0}{z_{(a)}}-\frac{(u_{(a)}^a+z_{(a)}\delta^a_0)(u_{(a)A}+z_{(a)}\delta_A^0)}{z_{(a)}(z_{(a)}+u^a\delta_a^0)} \,,
\end{align}
where $z_{(a)} = \sqrt{u_{(a)}^2}$. The boost operator satisfies the properties:
$B^{a}_{(a)A}B^{bA}_{(a)}=\eta^{ab}$ and $B^{a}_{(a)0}=u_{(a)}^a/z_{(a)}$.

To build the EFT description of dynamical tides, we start by modifying the Lagrangian~\eqref{eq_Lag_dytides_1} by 
introducing the conjugate momenta $P_{\mu\nu}$ with respect to the quadrupole moment, that is,
\begin{align}
    P_{(a)\mu\nu}=\frac{1}{c}\frac{\partial \mathcal{L}}{\partial({\dd Q_{(a)}^{\mu\nu}}/{\dd \tau})}=\frac{c}{2\lambda \omega_f^2 z_{(a)}} \frac{\dd Q_{(a)\mu\nu}}{\dd\tau}\,.
\end{align}
%
%
%
%
The new Lagrangian (in the Routhian form) obtained after making a Legendre transformation is given by
\begin{align}
\mathcal{L}_{{\rm DT} (a)}= c P_{(a)\mu\nu} \frac{\dd Q_{(a)}^{\mu\nu}}{\dd\tau}\underbrace{-
z_{(a)} \left[ \lambda_{(a)} \omega_{f(a)}^2 P_{(a)}^{\mu\nu}P_{(a)\mu\nu} + \frac{1}{4\lambda} Q_{(a)}^{\mu\nu}Q_{(a)\mu\nu} \right] -\frac{z_{(a)}}{2} E_{\mu\nu}Q_{(a)}^{\mu\nu}   \phantom{\Bigg|_{\frac{a}{b}}}  }_{-\text{Routhian}} \,.
\label{eq_L_DT_intermediate}
\end{align}
Here the advantage of working with $P_{\mu\nu}$, is that the new Lagrangian will depend only linearly on the complicated covariant derivative of the quadrupole moment tensor $Q_{\mu\nu}$~\cite{Steinhoff:2016rfi}. 
The supplementary condition for the dynamical degrees of freedom~\eqref{eq_Q_guage} in the rest frame of the star becomes \cite{Gupta:2020lnv}:
\begin{align}
Q_{(a)}^{A0}=0\,, \quad \text{and}\quad P_{(a)}^{A0}=0\,,
\end{align}
where, we now explicitly see that $Q^{AB}_{(a)}$ and $P^{AB}_{(a)}$ are spatial tensors that encode only the physical degrees of freedom. Thus, hereafter, we write the spatial tensor $Q_{(a)}^{AB}\delta_A^i\delta_B^j = \bm{Q}_{(a)}^{ij}$ and $P_{(a)}^{AB}\delta_A^i\delta_B^j = \bm{P}_{(a)}^{ij}$.

We can obtain the action for dynamical tides, written explicitly in terms of the physical degrees of freedom, $\bm{Q}_{(a)}^{ij}$ and $\bm{P}_{(a)}^{ij}$, by bringing the dynamical variables to the 
rest frame of each body by using the boost operator~\eqref{eq_boostopt} on the various terms in the Lagrangian~\eqref{eq_L_DT_intermediate}.
This gives us the effective point-particle (``pp'') action
\begin{align}\label{eq_action_pp}
S_{\text{pp}} = \sum_{a=1,2} \int \frac{\dd \tau}{c} \left[ - m_{(a)} z_{(a)} c^2 + {\cal L}_{{\rm FD}(a)} + {\cal L}_{{\rm MQ}(a)} + {\cal L}_{\rm{EQ}(a)} \right]\,.
\end{align}
The first term is simply the action for a point particle, while the remaining terms originate from the Lagrangian~\eqref{eq_L_DT_intermediate} as follows. 
The first term in Eq.~\eqref{eq_L_DT_intermediate} gives rise to,
\begin{align}
\mathcal{L}_{{\rm FD}(a)} &=
\bm{P}_{(a)}^{ij}\dot{\bm{Q}}_{(a)}^{ij}
+ c \left[ 
- u_{(a)}^\mu \omega_\mu^{ij} \left( \frac{\bm{S}_{Q(a)}^{ij}}{2} - \frac{\bm{S}_{Q(a)}^{ik} u^k_{(a)} u_{(a)}^j}{z_{(a)}(z_{(a)}+u_{(a)}^a \delta_a^0)} \right)
- u_{(a)}^\mu \omega_\mu^{ai}\delta_a^0 \bm{S}_{Q(a)}^{ij} \frac{u^j_{(a)}}{z_{(a)}}
\right.
\nonumber \\
&\qquad\qquad\qquad\quad\,\,
\left. 
+ \frac{\bm{S}_{Q(a)}^{ij}u^i_{(a)}}{z_{(a)}(z_{(a)}+u_{(a)}^a \delta_a^0)}\frac{\dd u^j_{(a)}}{\dd \tau}
\right] \,,
\end{align}
which describes frame-dragging (``FD'') effects on the quadrupole moment of each binary component. Here, we introduced the ``tidal spin'' tensor
\begin{equation}
\bm{S}_{Q(a)}^{ij} = 2 \, (\bm{Q}_{(a)}^{ki}\bm{P}_{(a)}^{jk}-\bm{Q}_{(a)}^{kj}\bm{P}_{(a)}^{ik}) \,,
\label{eq_spin_tensor}
\end{equation}
which describes the angular momentum of the dynamical quadrupole moment.
The second term in Eq.~\eqref{eq_L_DT_intermediate} yields,
\begin{align}
\mathcal{L}_{{\rm MQ}(a)} = -z_{(a)} \left[ \lambda_{(a)} \omega_{f(a)}^2 \bm{P}_{(a)}^{ij}\bm{P}_{(a)}^{ij} + \frac{1}{4\lambda_{(a)}} \bm{Q}_{(a)}^{ij}\bm{Q}_{(a)}^{ij} \right]
= -z_{(a)} \bm{M}_{Q(a)} \,.
\label{eq_L_MQ}
\end{align}
This term governs the dynamics of the quadrupole moment, which, by the second equality, can be described a time-dependent effective mass term for quadrupole moment (``MQ'').
The second equality is valid because all dependence on the gravitational field comes only through the redshift $z_{(a)}$. Thus now $\mathcal{L}_{{\rm MQ}(a)}$ becomes similar to the point mass terms [first term in Eq.~\eqref{eq_action_pp}], where the mass $\bm{M}_{Q(a)}$ is now time dependent.
Finally, the last term in Eq.~\eqref{eq_L_DT_intermediate} results in,
\begin{align}
\mathcal{L}_{{\rm EQ}(a)} =
- \frac{z_{(a)}}{2} \bm{E}_{(a)}^{ij}\bm{Q}_{(a)}^{ij} \,.
\end{align}
This term acts as a driving source for the quadrupole moment's dynamics and is induced on each of the binary components by the gravitoelectric tidal field $\bm{E}_{(a)}^{ij}=B^a_{(a)i}B^b_{(a)j}  e^\mu_{~a} e^\nu_{~b} E_{\mu\nu}$ of its companion.

In these equations, the indices contracted to the physical degrees of freedom, $\bm{Q}_{(a)}^{ij}$ and $\bm{P}_{(a)}^{ij}$, are understood to be in the rest frame of each star. A derivation of Eq.~\eqref{eq_action_pp} can be found in Ref.~\cite{Steinhoff:2016rfi}.
\section{Computational algorithm}
\label{sec_Routine}

Having obtained the effective point-particle action which includes dynamical gravitoelectric quadrupolar dynamical tides, we can now proceed to compute the effective two-body potential. 
In this section, we present the computational algorithm to perform this calculation. This potential will then be used in the next section to obtain the effective two-body Hamiltonian.

The dynamics of the gravitational field $g_{\mu\nu}$ is given by the Einstein-Hilbert action along with a harmonic gauge fixing term in $d+1$ spacetime dimensions,
\begin{align}\label{eq_action_EH}
S_{\text{EH}} = -\frac{c^4}{16 \pi G_d} \int \dd^{d+1} x~\sqrt{g} ~R + \frac{c^4}{32 \pi G_d } \int \dd^{d+1} x~\sqrt{g} ~g_{\mu\nu}~\Gamma^\mu \Gamma^\nu \, ,
\end{align}
where $\Gamma^\mu=\Gamma^{\mu}_{\rho\sigma}g^{\rho\sigma}$, 
$\Gamma^{\mu}_{\rho\sigma}$ is the Christoffel symbol, $R$ is the Ricci scalar, and $g$ is the metric determinant.
We work with the gravitational constant in $(d+1)$ spacetime dimensions written as $G_d = (\sqrt{4 \pi \exp(\gamma_{\rm E})} \, R_0 )^{d-3} \, G_{N}$. We express $G_d$ in this particular form because later on we will employ the 
modified minimal subtraction scheme~\cite{Collins:1984xc}, and hence the appearance of the $4 \pi$, the Euler-Mascheroni constant $\gamma_{\rm E}$, and the (arbitrary) lenghtscale $R_0$.

Since we are interested in the conservative dynamics of the system, we decompose the metric as $g_{\mu\nu}=\eta_{\mu\nu}+H_{\mu\nu}$, where $H_{\mu\nu}$ is the potential graviton is obtained after implementing the method of regions \cite{Beneke:1997zp} as done in Ref.~\cite{Goldberger:2004jt}.
We then decompose the metric using the standard Kaluza-Klein parametrization where the 10 degrees of freedom of $H_{\mu\nu}$ are encoded in three fields: a scalar $\bm{\phi}$, a three-dimensional vector $\bm{A}$ and a three-dimensional symmetric rank two tensor $\bm{\sigma}$~\cite{Kol:2007bc,Kol:2007rx}. In this parametrization, we write the metric as
\begin{equation}
g_{\mu\nu} = 
\begin{pmatrix}
e^{2\bm{\phi}/c^2} \,\,\, & -e^{2\bm{\phi}/c^2} {\bm{A}_j}/{c^2}\\
-e^{2\bm{\phi}/c^2} {\bm{A}_i}/{c^2} \,\,\,\,\,\, & -e^{-2\bm{\phi}/((d-2)c^2)}\bm{\gamma}_{ij}+e^{2\bm{\phi}/c^2} {\bm{A}_i}{\bm{A}_j}/{c^4}  
\end{pmatrix}\,,
\quad {\rm with} \quad
\bm{\gamma}_{ij}=\bm{\delta}_{ij}+\bm{\sigma}_{ij}/c^2 \,.
\end{equation}

We can now obtain the effective action for the binary by integrating out the gravitational degrees of freedom as follows,
\begin{align}
\label{eq_path_integral}
\exp \left[{\ii \, \int \dd t ~ \mathcal{L}_{\text{eff}}}\right] = \int \DD \bm{\phi} \, \DD \bm{A}_i \, \DD \bm{\sigma}_{ij}\, \exp[\ii \, (S_{\text{EH}}+S_{\text{pp}})] \, ,
\end{align}
where the Einstein-Hilbert action is given by Eq.~\eqref{eq_action_EH} and the point-particle action is given by Eq.~\eqref{eq_action_pp}.
To perform this integration, it is convenient to decompose the effective Lagrangian ${\cal L}_{\rm eff}$ as 
\begin{equation}
\mathcal{L}_{\rm eff} =  \mathcal{K}_{\rm eff} - \mathcal{V}_{\rm eff} \,,
\end{equation}
where $\mathcal{K}_{\rm eff}$ is an effective kinetic term, which does not dependent on any integration of potential graviton (i.e., it does not depend on integration of ${\bm \phi}$, ${\bm A}$, and ${\bm \sigma}$). We can compute $\mathcal{K}_{\rm eff}$ directly up to the required PN order. Explicitly, we decompose $\mathcal{K}_{\rm eff}$ in a point-particle, a frame-dragging, and a ``quadrupole mass'' contribution, i.e., ${\cal K}_{\rm eff} = {\cal K}_{\rm pp} + {\cal K}_{\rm FD} + {\cal K}_{\rm MQ}$,
\begin{subequations}
\label{eq_K_all}
\begin{align}\label{eq_Kpp}
\mathcal{K}_{\rm pp} =& \sum_{a=1,2} m_{(a)} 
\left[
\frac{1}{2}  \bm{v}_a^2 + \frac{1}{8}  \bm{v}_{(a)}^4 \left(\frac{1}{c^2}\right)  + \frac{1}{16}  \bm{v}_{(a)}^6 \left(\frac{1}{c^4}\right)
\right]
+ \mathcal{O}\left(\frac{1}{c^6}\right)\,, \\
\label{eq_KSO}
\mathcal{K}_{\rm FD} =& \sum_{a=1,2} 
\left\{ \bm{P}_{(a)}^{ij}\dot{\bm{Q}}_{(a)}^{ij} + \bm{S}_{Q(a)}^{ij}\bm{v}_{(a)}^i\bm{a}_{(a)}^j 
\left[
\frac{1}{2} \left(\frac{1}{c^2}\right)+ \frac{3}{8}  \bm{v}_{(a)}^2 \left(\frac{1}{c^4}\right) 
\right] \right\} + \mathcal{O}\left(\frac{1}{c^{6}}\right) \,, \\
\label{eq_KMQ}
\mathcal{K}_{\rm MQ} =& \sum_{a=1,2} \bm{M}_{(a)} 
\left[ 
1 +\frac{1}{2}  \bm{v}_a^2 \left(\frac{1}{c^2}\right)  + \frac{1}{8}  \bm{v}_{(a)}^4 \left(\frac{1}{c^4}\right)   \right]
+ \mathcal{O}\left(\frac{1}{c^6}\right)\,.
\end{align}
\end{subequations}

The terms that are obtained after performing the integral given in equation \eqref{eq_path_integral} are collectively denoted by the potential $\mathcal{V}_{\rm eff}$. These terms are computed by summing over the connected Feynman diagrams without graviton loops, as shown below,
\begin{align}
\mathcal{V}_{\text{eff}} = \ii \,
\lim_{d\rightarrow 3} 
\int \frac{\dd^d \bm{p}}{(2\pi)^d}~
e^{\ii \, \bm{p}\cdot (\bm{x}_{(1)}-\bm{x}_{(2)})}
\, 		
\parbox{25mm}{
	\begin{tikzpicture}[line width=1 pt,node distance=0.4 cm and 0.4 cm]
	\coordinate[label=left: ] (v1);
	\coordinate[right = of v1] (v2);
	\coordinate[right = of v2] (v3);
	\coordinate[right = of v3] (v4);
	\coordinate[right = of v4, label=right: \tiny$(2)$] (v5);
	\coordinate[below = of v1] (v6);
	\coordinate[below = of v6, label=left: ] (v7);
	\coordinate[right = of v7] (v8);
	\coordinate[right = of v8] (v9);
	\coordinate[right = of v9] (v10);
	\coordinate[right = of v10, label=right: \tiny$(1)$] (v11);
	\fill[black!25!white] (v8) rectangle (v4);
	\draw[fermionnoarrow] (v1) -- (v5);
	\draw[fermionnoarrow] (v7) -- (v11);
	\end{tikzpicture}
}
\, ,
\label{eq:effective_lagrangian}
\end{align}
where $\bm{p}$ is the linear momentum transferred between the two bodies. 
To calculate~\eqref{eq:effective_lagrangian}, we first generate all the topologies that correspond to graviton exchanges between the worldlines of the two compact objects. 
There is one topology at tree-level ($G_N$), two topologies at one-loop ($G_N^2$), and nine topologies at two-loop ($G_N^3$).
We then insert these topologies with the Kaluza-Klein fields $\bm{\phi}$, $\bm{A}$ and $\bm{\sigma}$. The number of diagrams\footnote{The diagrams which can be
obtained from the change in the label $1\leftrightarrow 2$, are not counted as separate diagrams.} appearing in the point-particle sector is given in Table~\ref{tbl_no_diag_non_spinning}, whereas that in the tidal sector are given in Tables~\ref{tbl_no_diag_tidalEQ},~\ref{tbl_no_diag_tidalFD}~and \ref{tbl_no_diag_tidalMQ}.
\begin{table}
	\begin{subtable}[H]{0.49\textwidth}
    \centering
		\begin{tabular}{|c|c|c|c|}
			\hline
			Order                & Diagrams            & Loops       & Diagrams  \\ \hline
			0PN                  & 1                   & 0      & 1   \\ \hline
			\multirow{2}{*}{1PN} & \multirow{2}{*}{4}  & 0    & 3 \\ \cline{3-4} 
			&                     & 1  & 1 \\ \hline
			\multirow{3}{*}{2PN} & \multirow{3}{*}{21} & 0    & 6 \\ \cline{3-4} 
			&                     & 1 & 10 \\ \cline{3-4} 
			&                     & 2  & 5 \\ \hline
		\end{tabular}
		\caption{Point particle sector}
		\label{tbl_no_diag_non_spinning}
	\end{subtable}
	\begin{subtable}[H]{0.49\textwidth}
    \centering
		\begin{tabular}{|c|c|c|c|}
			\hline
			Order                & Diagrams            & Loops      & Diagrams  \\ \hline
			0PN                  & 1                  & 0      &  1  \\ \hline
			\multirow{2}{*}{1PN} & \multirow{2}{*}{4}  & 0    & 3 \\ \cline{3-4} 
			&                     & 1  &  1 \\ \hline
			\multirow{3}{*}{2PN} & \multirow{3}{*}{26} & 0   & 6 \\ \cline{3-4} 
			&                     & 1 & 12  \\ \cline{3-4} 
			&                     & 2  &  7 \\ \hline
		\end{tabular}
		\caption{EQ sector}
		\label{tbl_no_diag_tidalEQ}
	\end{subtable}
	
	\vspace{0.5cm}
	\begin{subtable}[H]{0.49\textwidth}
    \centering
		\begin{tabular}{|c|c|c|c|}
			\hline
			Order                & Diagrams            & Loops      & Diagrams  \\ \hline
			1PN                  & 2                   & 0      &  2  \\ \hline
			\multirow{2}{*}{2PN} & \multirow{2}{*}{13}  & 0    & 5 \\ \cline{3-4} 
			&                     & 1  & 8 \\ \hline
		\end{tabular}
		\caption{FD sector}
		\label{tbl_no_diag_tidalFD}
	\end{subtable}
    \begin{subtable}[H]{0.49\textwidth}
    \centering
		\begin{tabular}{|c|c|c|c|}
			\hline
			Order                & Diagrams            & Loops      & Diagrams  \\ \hline
			1PN                  & 1                   & 0      &  1  \\ \hline	
			\multirow{2}{*}{2PN} & \multirow{2}{*}{4}  & 0    & 3 \\ \cline{3-4} 
			&                     & 1  & 1 \\ \hline
		\end{tabular}
		\caption{MQ sector}
		\label{tbl_no_diag_tidalMQ}
	\end{subtable}
	\label{tbl_no_diag}
	\caption{Number of Feynman diagrams contributing different sectors.
 }
\end{table}
We then compute these Feynman diagrams by means of an in-house code that uses tools from \texttt{EFTofPNG}~\cite{Levi:2017kzq} and \texttt{xTensor}~\cite{xAct}, for the tensor algebra manipulation, and \texttt{LiteRED} \cite{Lee:2013mka}, for the integration-by-parts reduction. 
This reduction recasts the Feynamn diagrams in terms of two point massless master integrals~\cite{Foffa:2016rgu} as shown in Fig.~\ref{fig_relation_bet_diags}.
\begin{figure}[H]
	\centering
	\begin{tikzpicture}[line width=1 pt, scale=0.4]
	\begin{scope}[shift={(-7,0)}]
	\filldraw[color=gray!40, fill=gray!40, thick](0,0) rectangle (3,3);
	\draw (-1.5,0)--(4.5,0);
	\draw (-1.5,3)--(4.5,3);
	\node at (1.5,6.5) {Gravity};	
	\node at (1.5,5.3) {Diagrams};	
	\end{scope}
	\begin{scope}[shift={(0,0)}]
	\node at (0,6) {$\longleftrightarrow$};
	\node at (0,1.5) {$\equiv$};
	\end{scope}
	\begin{scope}[shift={(5,0)}]
	\filldraw[color=gray!40, fill=gray!40, thick](0,1.5) circle (1.5);
	\draw (0,3)--(0,4);
	\draw (0,0)--(0,-1);	
	\node at (0,6.5) {Multi-loop};				
	\node at (0,5.3) {Diagrams};				
	\end{scope}		
	\end{tikzpicture}
	\caption{The diagrammatic correspondence between the four-point EFT-Gravity graphs
and the two-point quantum-field-theory (QFT) graphs.}
	\label{fig_relation_bet_diags}
\end{figure}
Once the exact expressions for the master integrals are substituted, we perform a Fourier transform to obtain the position-space effective potential ${\cal V}_{\rm eff}$. The details of the algorithm and the expressions for the master integrals up to two loops can be found in Ref.~\cite{Mandal:2022nty}.

After carrying out all these steps, the effective potential can be decomposed into
a point-particle and a dynamical tide contribution, i.e., ${\cal V}_{\rm eff} = {\cal V}_{\rm pp} + {\cal V}_{\rm DT}$, where
\begin{subequations}
\begin{align}
\mathcal{V}_{\rm pp} &= \mathcal{V}_{\rm N} + \left(\frac{1}{c^2}\right) \mathcal{V}_{\rm 1PN} + \left(\frac{1}{c^4}\right) \mathcal{V}_{\rm 2PN} + \mathcal{O}\left(\frac{1}{c^6}\right)\, , \\
\mathcal{V}_{\rm DT} &= \sum_{n = 0}^{2} \left(\frac{1}{c^2}\right)^n 
\left(
{\cal V}^{\rm EQ}_{n{\rm PN}}
+ {\cal V}^{\rm FD}_{n{\rm PN}}
+ {\cal V}^{\rm MQ}_{n{\rm PN}}
\right) 
+ \mathcal{O}\left(\frac{1}{c^6}\right) \,,
\end{align}
\end{subequations}
and we remark that ${\cal V}_{\rm DT}$ has contributions due to the driving source, the ``quadrupole-mass'' and the frame-dragging terms.

The potential ${\cal V}_{\rm eff}$ is now a function of the dynamical variables $\bm{x}_{(a)}$, $\bm{Q}_{(a)}$, and $\bm{S}_{Q(a)}$ and $\bm{M}_{Q(a)}$ through Eqs.~\eqref{eq_spin_tensor} and~\eqref{eq_L_MQ} as well, with higher order time derivatives included.
The first and higher-order time derivatives of $\bm{Q}_{(a)}$, $\bm{S}_{Q(a)}$, and $\bm{M}_{Q(a)}$ can be removed using integration by parts, while second and higher-order time derivatives of $\bm{x}_{(a)}$ 
are removed using a coordinate transformation $\bm{x}_{(a)} \rightarrow \bm{x}_{(a)} + \delta \bm{x}_{(a)}$. 
This coordinate transformation changes the Lagrangian as
\begin{align}
\delta \mathcal{L} = 
\frac{\delta \mathcal{L}}{\delta \bm{x}_{(a)}^i} \, \delta \bm{x}_{(a)}^i + \mathcal{O} (\delta \bm{x}_{(a)}^2) \,,
\end{align}
where $\delta \bm{x}_{(a)}$ is chosen such that it removes the undesirable terms from our final Lagrangian.
In our case, since we work up to 2PN order, the process of removing the higher order time derivatives using a coordinate transformation is equivalent to the substitution of the equation of motion for the acceleration $\bm{a}_{(a)}$ and its higher order time derivatives back into the Lagrangian.
In the end, this procedure distils the Lagrangian into a final form which depends only on $\bm{x}_{(a)}$, $\bm{v}_{(a)}$, and $\bm{Q}_{(a)}$.
\section{The effective Hamiltonian for dynamical tides}
\label{sec_results}

In this section, we present the result of the effective two-body Hamiltonian with dynamical gravitoelectric tides. This Hamiltonian $\mathcal{H}$ is computed from the Lagrangian obtained in the previous section using a Legendre transformation
\begin{align}\label{eq_ham_def}
\mathcal{H}(\bm{x},\bm{p},\bm{Q})= \sum_{a=1,2}( \, \bm{p}_{(a)}^i {\bm v}_{(a)}^i  +\bm{P}_{(a)}^{ij}\dot{\bm{Q}}_{(a)}^{ij} \,) - \mathcal{L}(\bm{x},{\bm v},\bm{Q}) \ .
\end{align}
To express this Hamiltonian in a compact form we introduce a few variables. 
The total mass of the binary is denoted by $M=m_{(1)}+m_{(2)}$, the reduced mass by $\mu=m_{(1)}m_{(2)}/M$, the mass ratio by $q=m_{(1)}/m_{(2)}$, the symmetric mass ratio by $\nu=\mu/M$, and the antisymmetric mass ratio $\delta=(m_{(1)}-m_{(2)})/M$, which are related to each other by,
\begin{align}
\nu=\frac{m_{(1)} m_{(2)}}{M^2}=\frac{\mu}{M}=\frac{q}{(1+q)^2} = \frac{(1-\delta^2)}{4} \,.
\end{align}
We express the results in the center-of-mass (COM) frame of reference and define the momentum in the COM frame as $\bm{p} \equiv \bm{p}_{(1)}=-\bm{p}_{(2)}$. In the COM frame, the orbital angular momentum is defined as $\bm{L} = \bm{r}\times \bm{p}$. Hence, we can write $p^2 = p_r^2 + L^2 / r^2$, where $p_r=\bm{p} \cdot \bm{n}$, $p = |\bm{p}|$ and $L = |\bm{L}|$.
We rescale all the variables to express the Hamiltonian in terms of dimensionless quantities, which we denote by a tilde as follows
\begin{align}
&\widetilde{\bm{p}}=\frac{1}{c}\frac{\bm{p}}{\mu} \, ,\quad
\widetilde{\bm{r}}=\frac{c^2}{G_N} \frac{\bm{r}}{M} \, ,\quad
\widetilde{\bm{L}}=\frac{c}{G_N} \frac{\bm{L}}{M \mu} \, ,\quad 
\widetilde{\mathcal{H}}=\frac{1}{c^2} \frac{\mathcal{H}}{\mu}\,,\nonumber \\
\label{eq_rescale}
&\widetilde{\bm{Q}}_{(a)}=\frac{c^4}{G_N^2} \frac{\bm{Q}_{(a)}}{M^2 \mu} \, ,\quad
\widetilde{\bm{S}}_{Q(a)}=\frac{c}{G_N} \frac{\bm{S}_{Q(a)}}{M \mu} \,, \quad \text{and} \quad
\widetilde{\bm{M}}_{Q(a)}=\frac{1}{c^2}\frac{\bm{M}_{Q(a)}}{\mu} \,.
\end{align}
The total EFT Hamiltonian in the dimensionless parameters is given by
\begin{align}\label{eq_ham_dy}
\widetilde{\mathcal{H}} = \widetilde{\mathcal{H}}_{\rm pp} + \widetilde{\mathcal{H}}_{\rm DT}\ ,
\end{align}
where
\begin{subequations}
\begin{align}\label{eq_Ham_pp}
\widetilde{\mathcal{H}}_{\rm pp} &= \widetilde{\mathcal{H}}_{\text{0PN}}+\left(\frac{1}{c^2}\right)\widetilde{\mathcal{H}}_{\text{1PN}}+\left(\frac{1}{c^4}\right)\widetilde{\mathcal{H}}_{\text{2PN}}
+\mathcal{O}\left(\frac{1}{c^6}\right) \,, \\
\widetilde{{\cal H}}_{\rm DT} &= \sum_{n = 0}^{2}
\left(\frac{1}{c^2}\right)^n \, 
\left(
\widetilde{{\cal H}}^{\rm EQ}_{n{\rm PN}}
+ \widetilde{{\cal H}}^{\rm FD}_{n{\rm PN}}
+ \widetilde{{\cal H}}^{\rm MQ}_{n{\rm PN}}
\right)
+\mathcal{O}\left(\frac{1}{c^6}\right) \,.
\end{align}
\end{subequations}
The point particle Hamiltonian till 2PN is presented in the same gauge in Appendix C.1 of Ref.~\cite{Mandal:2022nty}, and the tidal sector of the Hamiltonian is known till 1PN~\cite{Steinhoff:2016rfi}. The new result for the tidal Hamiltonian at 2PN is presented here for the first time to the best of our knowledge.

The leading order contribution at 0PN order is given as
\begin{subequations}
\label{eq:old_0PN_H}
\begin{align}
\widetilde{\mathcal{H}}^{\rm EQ}_{\text{0PN}} &=
\left( \widetilde{\bm{Q}}^{ij}_{(1)}\widetilde{\bm{r}}^{i} \widetilde{\bm{r}}^{j} \right)
\left(
-\frac{3 \nu }{2\widetilde{r}^5}-\frac{1}{q}\frac{3 \nu }{2\widetilde{r}^5}
\right)
+ (1\leftrightarrow 2) \,, \\
\widetilde{\mathcal{H}}^{\rm FD}_{\text{0PN}} &= 0  \,,\\
\widetilde{\mathcal{H}}^{\rm MQ}_{\text{0PN}} &= \widetilde{\bm{M}}_{Q(1)} +(1\leftrightarrow 2) \,.
\end{align}
\end{subequations}
The next-to-leading order terms at 1PN order are
\begin{subequations}
\label{eq:old_1PN_H}
\begin{align}    
\widetilde{\mathcal{H}}^{\rm EQ}_{\text{1PN}} &= 
\left( \widetilde{\bm{Q}}^{ij}_{(1)}\widetilde{\bm{r}}^{i} \widetilde{\bm{r}}^{j} \right)
\, \left\{
\widetilde{L}^2 \left(-\frac{3 \nu ^2}{4\widetilde{r}^7}-\frac{3 \nu }{\widetilde{r}^7}\right)+\widetilde{p}_r^2 \left(\frac{9 \nu }{4\widetilde{r}^5}-\frac{3 \nu ^2}{\widetilde{r}^5}\right)+\frac{15 \nu }{4\widetilde{r}^6} \right. \nonumber\\
&\quad \left. 
+\frac{1}{q} \left[\widetilde{L}^2 \left(-\frac{3 \nu ^2}{4\widetilde{r}^7}-\frac{9 \nu }{4\widetilde{r}^7}\right)+\widetilde{p}_r^2 \left(\frac{3 \nu }{4\widetilde{r}^5}-\frac{3 \nu ^2}{\widetilde{r}^5}\right)+\frac{6 \nu }{\widetilde{r}^6}\right]
\right\}  
+ 
\left(\widetilde{\bm{Q}}^{ij}_{(1)}\widetilde{\bm{L}}^i\widetilde{\bm{L}}^j\right) 
\left(
-\frac{3 \nu }{2\widetilde{r}^5}-\frac{1}{q}\frac{3 \nu }{2\widetilde{r}^5} 
\right)
\nonumber\\
&\quad 
+ \left(\widetilde{\bm{Q}}^{ij}_{(1)}\widetilde{\bm{r}}^i\widetilde{\bm{L}}^j \right) \, \widetilde{p}_r 
\left[ 
\frac{3 \nu ^2}{2\widetilde{r}^5}+ \frac{1}{q} \left(\frac{3 \nu ^2}{2\widetilde{r}^5}+\frac{3 \nu }{2\widetilde{r}^5}\right)
\right]
+(1\leftrightarrow 2) \,, \\
\widetilde{\mathcal{H}}^{\rm FD}_{\rm 1PN} &=
\left(\widetilde{\bm{S}}_{Q(1)}\cdot \widetilde{\bm{L}}\right) \left( \frac{2 \nu }{\widetilde{r}^3}+\frac{1}{q}\frac{3 \nu}{2 \widetilde{r}^3} \right) 
+ (1\leftrightarrow 2) \,,\\
\widetilde{\mathcal{H}}^{\rm MQ}_{\text{1PN}} &= \widetilde{\bm{M}}_{Q(1)}\left[
-\frac{\nu }{\widetilde{r}}
+ \frac{1}{q}
\left(
-\widetilde{L}^2\frac{ \nu }{2\widetilde{r}^2}-\widetilde{p}_r^2\frac{\nu}{2}-\frac{\nu }{\widetilde{r}}
\right)
\right]
+ (1\leftrightarrow 2) \,.
\end{align}
\end{subequations}
And finally the novel contributions of the next-to-next-to-leading order terms at 2PN order are
\begin{subequations}
\label{eq:new_2PN_H}
\begin{align} 
\widetilde{\mathcal{H}}^{\rm EQ}_{\rm 2PN} &= 
\left( \widetilde{\bm{Q}}^{ij}_{(1)}\widetilde{\bm{r}}^{\,i}\widetilde{\bm{r}}^{\,j} \right)
\left\{
\widetilde{L}^4 \left(-\frac{9 \nu ^3}{16\widetilde{r}^9}-\frac{63 \nu ^2}{16\widetilde{r}^9}\right)+\widetilde{L}^2 \left[\widetilde{p}_r^2 \left(-\frac{15 \nu ^3}{8\widetilde{r}^7}-\frac{45 \nu ^2}{4\widetilde{r}^7}+\frac{15 \nu }{2\widetilde{r}^7}\right)+\frac{225 \nu ^2}{16\widetilde{r}^8}+\frac{149 \nu }{16\widetilde{r}^8}\right] \right. \nonumber\\
&\quad +\widetilde{p}_r^2 \left(\frac{75 \nu ^2}{8\widetilde{r}^6}-\frac{27 \nu }{4\widetilde{r}^6}\right) +\widetilde{p}_r^4 \left(-\frac{9 \nu ^3}{2\widetilde{r}^5}+\frac{9 \nu ^2}{\widetilde{r}^5}-\frac{45 \nu }{16\widetilde{r}^5}\right) -\frac{183 \nu ^2}{28\widetilde{r}^7}-\frac{285 \nu }{56\widetilde{r}^7} \nonumber\\
&\quad +\frac{1}{q} \Bigg[\widetilde{L}^4 \left(-\frac{9 \nu ^3}{16\widetilde{r}^9}-\frac{45 \nu ^2}{16\widetilde{r}^9}+\frac{15 \nu }{16\widetilde{r}^9}\right)+\widetilde{L}^2 \left(\widetilde{p}_r^2 \left(-\frac{15 \nu ^3}{8\widetilde{r}^7}-\frac{33 \nu ^2}{4\widetilde{r}^7}+\frac{3 \nu }{4\widetilde{r}^7}\right)+\frac{231 \nu ^2}{16\widetilde{r}^8}+\frac{27 \nu }{2\widetilde{r}^8}\right)\nonumber\\
&\quad \left. +\widetilde{p}_r^2 \left(\frac{63 \nu ^2}{4\widetilde{r}^6}-\frac{9 \nu }{2\widetilde{r}^6}\right)+\widetilde{p}_r^4 \left(-\frac{9 \nu ^3}{2\widetilde{r}^5}+\frac{9 \nu ^2}{2\widetilde{r}^5}-\frac{9 \nu }{16\widetilde{r}^5}\right)-\frac{183 \nu ^2}{28\widetilde{r}^7}-\frac{57 \nu }{4\widetilde{r}^7}\Bigg]
\right\}\nonumber\\
&\quad + \left(\widetilde{\bm{Q}}^{ij}_{(1)}\widetilde{\bm{L}}^i\widetilde{\bm{L}}^j\right)
\left\{
\widetilde{p}_r^2 \left(-\frac{3 \nu ^3}{8\widetilde{r}^5}-\frac{45 \nu ^2}{8\widetilde{r}^5}+\frac{15 \nu }{4\widetilde{r}^5}\right)+\widetilde{L}^2\left(-\frac{9  \nu ^2}{4\widetilde{r}^7}\right)+\frac{3 \nu ^2}{\widetilde{r}^6}+\frac{43 \nu }{4\widetilde{r}^6} \right. \nonumber\\
&\quad\left. +\frac{1}{q} \left[\widetilde{L}^2 \left(\frac{3 \nu }{4\widetilde{r}^7}-\frac{9 \nu ^2}{4\widetilde{r}^7}\right)+\widetilde{p}_r^2 \left(-\frac{3 \nu ^3}{8\widetilde{r}^5}-\frac{51 \nu ^2}{8\widetilde{r}^5}+\frac{3 \nu }{8\widetilde{r}^5}\right)+\frac{3 \nu ^2}{\widetilde{r}^6}+\frac{9 \nu }{\widetilde{r}^6}\right]
\right\} \nonumber\\
&\quad + \left( \widetilde{\bm{Q}}^{ij}_{(1)}\widetilde{\bm{r}}^{\,i} \widetilde{\bm{L}}^j \right) 
\widetilde{p}_r \left\{
\widetilde{L}^2 \left(\frac{9 \nu ^3}{8\widetilde{r}^7}+\frac{15 \nu ^2}{8\widetilde{r}^7}\right)+\widetilde{p}_r^2 \left(\frac{33 \nu ^3}{8\widetilde{r}^5}-\frac{39 \nu ^2}{8\widetilde{r}^5}\right)-\frac{15 \nu ^2}{4\widetilde{r}^6}-\frac{91 \nu }{4\widetilde{r}^6} \right. \nonumber\\
&\quad\left. +\frac{1}{q} \left[\widetilde{L}^2 \left(\frac{9 \nu ^3}{8\widetilde{r}^7}+\frac{21 \nu ^2}{8\widetilde{r}^7}-\frac{3 \nu }{8\widetilde{r}^7}\right)
+\widetilde{p}_r^2 \left(\frac{33 \nu ^3}{8\widetilde{r}^5}+\frac{15 \nu ^2}{8\widetilde{r}^5}-\frac{9 \nu }{8\widetilde{r}^5}\right) -\frac{27 \nu ^2}{4\widetilde{r}^6}-\frac{9 \nu }{\widetilde{r}^6}\right]
\right\}
\nonumber\\
&\quad+(1\leftrightarrow 2) \,, \\[10pt]
\widetilde{\mathcal{H}}^{\rm FD}_{\rm 2PN} &=
\left( \widetilde{\bm{S}}_{Q(1)}\cdot \widetilde{\bm{L}} \right)  \left\{ -\frac{5 \nu ^2}{4 \widetilde{r}^4}-\frac{7 \nu }{\widetilde{r}^4}+\widetilde{p}_r^2 \left(\frac{43 \nu ^2}{8\widetilde{r}^3}-\frac{2 \nu }{\widetilde{r}^3}\right) \right.
+\widetilde{L}^2 \left(\frac{13 \nu ^2}{8\widetilde{r}^5}+\frac{\nu }{\widetilde{r}^5}\right) \nonumber\\
&\quad \left. +\frac{1}{q}\left[ -\frac{5 \nu ^2}{4 \widetilde{r}^4}-\frac{5 \nu }{\widetilde{r}^4}+\widetilde{p}_r^2 \left(\frac{17 \nu ^2}{4 \widetilde{r}^3}-\frac{5 \nu }{8 \widetilde{r}^3}\right)+
\widetilde{L}^2 \left(\frac{5 \nu ^2}{4 \widetilde{r}^5}-\frac{5 \nu }{8 \widetilde{r}^5}\right)
\right]
\right\} + (1\leftrightarrow 2) \,,\\[10pt]
\widetilde{\mathcal{H}}^{\rm MQ}_{\text{2PN}} &= \widetilde{\bm{M}}_{Q(1)}
\left\{\widetilde{L}^4 \left(-\frac{3  \nu ^2}{8\widetilde{r}^4}\right)+\widetilde{L}^2 \left[\widetilde{p}_r^2\left(-\frac{3 \nu ^2}{4\widetilde{r}^2}\right)-\frac{2 \nu }{\widetilde{r}^3}\right] +\widetilde{p}_r^2 \left(-\frac{3 \nu }{2 r}\right) +\widetilde{p}_r^4 \left(-\frac{3 \nu ^2 }{8} \right) +\frac{3 \nu }{2\widetilde{r}^2} \right. \nonumber\\
&\quad \left. +\frac{1}{q} \left[\widetilde{L}^4 \left(\frac{3 \nu }{8\widetilde{r}^4}-\frac{3 \nu ^2}{4\widetilde{r}^4}\right)+\widetilde{L}^2 \left(\widetilde{p}_r^2 \left(\frac{3 \nu }{4\widetilde{r}^2}-\frac{3 \nu ^2}{2\widetilde{r}^2}\right)+\frac{3 \nu }{2\widetilde{r}^3}\right) +\widetilde{p}_r^{\,2} \, \frac{3 \nu}{2 r} + \widetilde{p}_r^4 \left(\frac{3 \nu }{8}-\frac{3 \nu ^2}{4}\right)+\frac{\nu }{2\widetilde{r}^2}\right]
\right\} \nonumber\\
&\quad +(1\leftrightarrow 2) \,.
\end{align}
\end{subequations}
Equations~\eqref{eq:new_2PN_H}, together with the previously known results~\eqref{eq:old_0PN_H} and~\eqref{eq:old_1PN_H}, complete the description of the conservative dynamics of gravitoelectric dynamical quadrupolar tidal interaction in a nonspinning compact binary at 2PN order.

We remark that Steinhoff et al.~\cite{Steinhoff:2016rfi}, who worked to 1PN order, had observed that $\widetilde{\mathcal{H}}^{\text{EQ}}$ and $\widetilde{\mathcal{H}}^{\text{FD}}$ are similar to the next-to-leading order spin-induced quadrupole and leading-order spin-orbit Hamiltonians~\cite{Levi:2015msa} respectively, upon applying certain replacements. Here, we have derived Eqs.~\eqref{eq:old_1PN_H}~and~\eqref{eq:new_2PN_H} from first principles, starting from the effective action~\eqref{eq_action_pp}.
As an additional confirmation of our result, we used the analogy identified in Ref.~\cite{Steinhoff:2016rfi}, to find a canonical transformation from the spin-induced quadrupole Hamiltonian up to next-to-next-to-leading order~\cite{Mandal:2022ufb} to $\widetilde{\mathcal{H}}^{\text{EQ}}$, and from the spin-orbit Hamiltonian up to next-to-leading order~\cite{Mandal:2022nty} to $\widetilde{\mathcal{H}}^{\text{FD}}$.
The effective Hamiltonian in a general reference frame is provided by us in the ancillary file \texttt{Hamiltonian-DT.m}.

\section{The adiabatic limit}
\label{sec_adiabatic}
In this section, we specialize our results to the limit of adiabatic tides, that is, we take the $\omega_f\rightarrow\infty$ limit of the Hamiltonian~\eqref{eq_ham_dy}. This eliminates the dependence of Hamiltonian on the variables $\bm{Q}_{(a)}$, $\bm{S}_{Q(a)}$ and $\bm{M}_{Q(a)}$, and hence simplifies our further calculations. We then compute the Poincar\'e algebra to validate the result of adiabatic Hamiltonian. Finally, we compute the binding energy and scattering angle using the adiabatic Hamiltonian and compare these against known results in the literature.


The adiabatic limit physically refers to the quadrupole mode being locked to the external tidal field induced by the binary companion. This is obtained by taking $\omega_f\rightarrow \infty$ in the Hamiltonian given in Eq.~\eqref{eq_ham_dy}. In this limit, the equation of motion of $\bm{Q}_{(a)}^{ij}$ is given by,
\begin{align}
  0= \frac{\partial \mathcal{H}}{\partial \bm{Q}_{(a)}^{ij}} &= -\frac{z_{(a)}}{2 \lambda_{(a)}} \bm{Q}_{(a)ij} - \frac{\partial \mathcal{H}_{\rm EQ}}{\partial \bm{Q}_{(a)}^{ij}} \,.
\end{align}
Here, to make the physical interpretation more clear, we define the tidal field in terms of the dynamic Hamiltonian itself, which leads to the following equation of motion for the $\bm{Q}_{(a)}^{ij}$,
\begin{align}
    \bm{Q}_{(a)}^{ij}=-\lambda_{(a)} \frac{2 }{z_{(a)}} \frac{\partial \mathcal{H}_{\rm EQ}}{\partial \bm{Q}_{(a)}^{ij}} = -\lambda_{(a)} \bm{E}_{(a)}^{ij}
\end{align}
We then compute the above equation of motion up to 2 PN order and substitute it in the Hamiltonian~\eqref{eq_ham_dy} to obtain the effective Hamiltonian for adiabatic tides.

We remark that we could also have derived the adiabatic Hamiltonian starting from the action~\eqref{eq_Lag_ad}. Following this route, we would have to do a three-loop computation to obtain the 2PN adiabatic effective Hamiltonian. 
However, one can easily show that all the Feynman diagrams appearing in the calculation would be factorizable due to the $E^2$ term. 
This means that all Feynman integrals that would appear in this calculation would also be factorizable into two-loop master integrals. This reveals an advantage of computing the adiabatic limit from the dynamical case: we have to compute one less loop integral to obtain the same result at a given PN order.

\subsection{The effective Hamiltonian for adiabatic tides}
Similar to the dynamical case, we first rescale all the variables to write the expressions for the Hamiltonian in terms of the dimensionless 
parameters given in Eq.~\eqref{eq_rescale} and we also introduce
\begin{align}\label{eq_lambda_rescaling}
\widetilde{\lambda}_{(a)} = \frac{c^{10}}{G_N^4}\frac{\lambda_{(a)}}{M^5} \,.
\end{align}
The adiabatic effective Hamiltonian in the dimensionless form can be written as\footnote{At leading order, adiabatic tides contributes at 5PN which can be easily seen writing the Hamiltonian in the form of the dimensional Love number as shown in Eq.~\eqref{eq_lambda_rescaling}. We show here the relative scaling with respect to the 5PN order contribution.}
\begin{align}\label{eq_ham_ad}
\widetilde{\mathcal{H}} = \widetilde{\mathcal{H}}_{\rm pp} + \widetilde{\mathcal{H}}_{\rm AT}\ ,
\end{align}
where
\begin{subequations}
\begin{align}
\widetilde{\mathcal{H}}_{\rm pp} &= \widetilde{\mathcal{H}}_{\text{0PN}}+\left(\frac{1}{c^2}\right)\widetilde{\mathcal{H}}_{\text{1PN}}+\left(\frac{1}{c^4}\right)\widetilde{\mathcal{H}}_{\text{2PN}}
+\mathcal{O}\left(\frac{1}{c^6}\right) \ , \\
\widetilde{\mathcal{H}}_{\rm AT} 
&= \widetilde{\mathcal{H}}^{\text{AT}}_{\text{0PN}}+\left(\frac{1}{c^{2}}\right)\widetilde{\mathcal{H}}^{\text{AT}}_{\text{1PN}}+\left(\frac{1}{c^{4}}\right)\widetilde{\mathcal{H}}^{\text{AT}}_{\text{2PN}}
+\mathcal{O}\left(\frac{1}{c^{6}}\right) \ , 
\end{align}
\end{subequations}
The point particle result is known and is also presented in the same gauge in Ref.~\cite{Mandal:2022nty}. The tidal sector of the Hamiltonian is given as
\begin{subequations}
\begin{align}
\widetilde{\mathcal{H}}^{\text{AT}}_{\text{0PN}} &= \widetilde{\lambda}_{(1)} \, 
\frac{1}{q} 
\left(-\frac{3}{2\widetilde{r}^6}\right)
+(1\leftrightarrow 2) \,,\\[10pt]
\widetilde{\mathcal{H}}^{\text{AT}}_{\text{1PN}} &= \widetilde{\lambda}_{(1)} \, \left\{
-\frac{3 \nu }{\widetilde{r}^7}
+\widetilde{p}_r^2 \left(\frac{15 \nu }{4\widetilde{r}^6}\right)
+\widetilde{L}^2 \left(-\frac{3 \nu }{4\widetilde{r}^8}\right) \right.\nonumber\\
&\left. \quad\quad\quad+\frac{1}{q}\left[
\frac{21}{2\widetilde{r}^7}-\frac{3 \nu }{\widetilde{r}^7}
+\widetilde{p}_r^2 \left(\frac{3}{4\widetilde{r}^6}-\frac{3 \nu }{2\widetilde{r}^6}\right)
+\widetilde{L}^2 \left(-\frac{3 \nu }{2\widetilde{r}^8}-\frac{15}{4\widetilde{r}^8}\right)
\right]
\right\}
+(1\leftrightarrow 2) \,,\\[10pt]
\widetilde{\mathcal{H}}^{\text{AT}}_{\text{2PN}} &= 
\widetilde{\lambda}_{(1)} \, 
\left\{
\frac{1215 \nu }{56\widetilde{r}^8}
+\widetilde{p}_r^2 \left(-\frac{15 \nu ^2}{\widetilde{r}^7}-\frac{18 \nu }{\widetilde{r}^7}\right)
+\widetilde{p}_r^4 \left(\frac{135 \nu ^2}{8\widetilde{r}^6}-\frac{135 \nu }{16\widetilde{r}^6}\right) 
\right.
\nonumber \\
&\quad \left.
+ \, \widetilde{L}^2 \left[\widetilde{p}_r^2 \left(\frac{135 \nu }{8\widetilde{r}^8}-\frac{9 \nu ^2}{4\widetilde{r}^8}\right)-\frac{3 \nu ^2}{4\widetilde{r}^9}-\frac{117 \nu }{8\widetilde{r}^9}\right]
+\widetilde{L}^4 \left(-\frac{9 \nu ^2}{8\widetilde{r}^{10}}-\frac{27 \nu }{16\widetilde{r}^{10}}\right) \right.
\nonumber \\
&\quad \left. 
+\frac{1}{q} \left[ \frac{21 \nu }{2\widetilde{r}^8}-\frac{165}{4\widetilde{r}^8}
+\widetilde{p}_r^2 \left(-\frac{15 \nu ^2}{\widetilde{r}^7}+\frac{81 \nu }{4\widetilde{r}^7}-\frac{27}{4\widetilde{r}^7}\right)
+\widetilde{p}_r^4 \left(\frac{81 \nu ^2}{16\widetilde{r}^6}+\frac{9 \nu }{4\widetilde{r}^6}-\frac{9}{16\widetilde{r}^6}\right) \right. \right.
\nonumber \\
&\quad\quad
\left. \left.
+ \, \widetilde{L}^2 \left(\widetilde{p}_r^2 \left(-\frac{45 \nu ^2}{8\widetilde{r}^8}-\frac{9 \nu }{2\widetilde{r}^8}+\frac{9}{8\widetilde{r}^8}\right)-\frac{3 \nu ^2}{4\widetilde{r}^9}+\frac{6 \nu }{\widetilde{r}^9}+\frac{135}{4\widetilde{r}^9}\right)
\right. \right.
\nonumber \\
&\quad\quad \left. \left.
+ \, \widetilde{L}^4 \left(-\frac{27 \nu ^2}{16\widetilde{r}^{10}}-\frac{27 \nu }{4\widetilde{r}^{10}}-\frac{45}{16\widetilde{r}^{10}}\right)
\right] \right\} + (1\leftrightarrow 2) \,.
\end{align}
\end{subequations}
This Hamiltonian in a general reference frame is provided by us in the ancillary file \texttt{Hamiltonian-AT.m}.
We were also able to find a canonical transformation from the Hamiltonian in the generic frame in Eq.~\eqref{eq_ham_ad} to the Hamiltonian found in Ref.~\cite{Henry:2019xhg}, which validates our result.

\subsection{The Poincar\'e algebra}
In this section, we validate the adiabatic effective Hamiltonian~\eqref{eq_ham_ad} by deriving the complete Poincar\'e algebra~\cite{Damour:2000kk,Levi:2016ofk}. 
This amounts to computing all the generators of the Poincar\'e algebra given by 
\begin{subequations}
\label{eq_full_Poincare}
\begin{align}
\{P^\mu,P^\nu\}&=0 \,,\\
\{P^\mu,J^{\rho\sigma}\}&=-\eta^{\mu\rho} P^\sigma + \eta^{\mu\sigma} P^\rho \,,\\
\{J^{\mu\nu},J^{\rho\sigma}\}&=-\eta^{\nu\rho} J^{\mu\sigma} +\eta^{\mu\rho} J^{\nu\sigma} +\eta^{\mu\sigma} J^{\rho\nu}-\eta^{\sigma\nu} J^{\mu\rho} \,,
\end{align}
\end{subequations}
where $P^\mu$ is the linear momentum and $J^{\mu\nu}$ is the angular momentum. These Poisson brackets can be decomposed into spatial and temporal parts. 
For later convenience, we separate them into two sets,
\begin{align}\label{eq_set1_poincare_algebra}
&\{\bm{P}^i,\mathcal{H}\}=0\,,\quad
\{\bm{J}^i,\mathcal{H}\}=0\,, \quad
\{\bm{J}^i,\bm{P}^j\}=\epsilon^{ijk}\bm{P}^k\,, \quad
\{\bm{J}^i,\bm{P}^j\}=\delta^{ij}\mathcal{H}\,, \quad
\{\bm{J}^i,\bm{J}^j\}=\epsilon^{ijk}\bm{J}^k \,,
\end{align}
and
\begin{align}
\label{eq_set2_poincare_algebra}
&\{\bm{J}^i,\bm{G}^j\}=\epsilon^{ijk}\bm{G}^k \,,\quad
\{\bm{G}^i,\bm{P}^j\}= \frac{1}{c^2}\delta^{ij}\mathcal{H} \,,\quad
\{\bm{G}^i,\mathcal{H}\}=\bm{P}^k \,,\quad
\{\bm{G}^i,\bm{G}^j\}=-\frac{1}{c^2}\epsilon^{ijk}\bm{J}^k \,,
\end{align}
where $\bm{P}^i$ are spatial components of the linear momentum, 
$\bm{J}^i$ are spatial components of the angular momentum, $\mathcal{H}$ is the Hamiltonian, and the boost generator is written as $\bm{K}^i=\bm{G}^i-t \bm{P}^i$, where $\bm{G}^i$ is the COM vector.

We begin by writing the linear momentum and angular momentum of the system as
\begin{align}
\bm{P}^i = \bm{p}_{(1)}^i+\bm{p}_{(2)}^i \,, \quad {\rm and} \quad
\bm{J}^i = \epsilon^{ijk}\bm{x}_{(1)}^j \bm{p}_{(1)}^j+\epsilon^{ijk}\bm{x}_{(2)}^j \bm{p}_{(2)}^k \,.
\end{align}
This ensures that the first set of Poisson brackets in Eq.~\eqref{eq_set1_poincare_algebra} is satisfied. Now our goal is to come up with an expression for the COM vector $\bm{G}^i$ such that the second set of Poisson brackets~\eqref{eq_set2_poincare_algebra} is also satisfied. 
To do so, we use the first two Poisson brackets in Eq.~\eqref{eq_set2_poincare_algebra}, i.e., $\{\bm{J}^i,\bm{G}^j\}=\epsilon^{ijk}\bm{G}^k $ and $\{\bm{G}^i,\bm{P}^j\}=(1/c^2)\delta^{ij}\mathcal{H}$, to make an ansatz for $\bm{G}$. This suggests that we should make the following ansatz for the COM vector,
\begin{align}
\bm{G}^i=  \left(\frac{1}{c^2}\right)\frac{\mathcal{H}}{2} (\bm{x}^i_{(1)}+\bm{x}^i_{(2)}) + h \bm{r}^i + \bm{Y}^i \,,
\end{align}
where $h$ is antisymmetric in $1\leftrightarrow 2$, while $\bm{Y}$ is symmetric in $1\leftrightarrow 2$.
We can now fix uniquely the above ansatz, in other words, we can determine $h$ and ${\bm Y}$, by using the third Poisson bracket of Eq.~\eqref{eq_set2_poincare_algebra}, i.e., $\{\bm{G}^i,\mathcal{H}\}=\bm{P}^k$.
Once we have the ansatz is uniquely fixed, we can check its validity by verifying that it does satisfy the last Poisson bracket in Eq.~\eqref{eq_set2_poincare_algebra}, i.e., $\{\bm{G}^i,\bm{G}^j\}=-(1/c^2)\epsilon^{ijk}\bm{J}^k$.
Following this procedure, we can determine uniquely $h$ and ${\bm Y}^{i}$ appearing in the COM vector ${\bm G}^{i}$, up to 2PN order:
\begin{align}
h &= \frac{m_{(1)}}{2}
+ \left( \frac{1}{c^2} \right) \left\{ \frac{\bm{p}_{(1)}^2}{4m_{(1)}} 
+  \lambda_{(1)} \left( -G^2 \frac{3  m_{(2)}^2}{4 r^6}  \right) \right\}
+ \left( \frac{1}{c^4} \right)  \left\{ G^2 \frac{m_{(1)}^2 m_{(2)}}{4 r^2}-\frac{\bm{p}_{(1)}^4}{16 m_{(1)}^3} 
\right.\nonumber\\
&\quad 
+ \left. \lambda_{(1)} \left[ \, \frac{3 G^3 m_{(2)}^2 \left(5 m_{(1)}+7 m_{(2)}\right)}{4 r^7} 
+ G^2 \left( -\frac{9 m_{(2)} (\bm{p}_{(1)}\cdot\bm{n}) (\bm{p}_{(2)}\cdot\bm{n})}{2 m_{(1)} r^6}+\frac{9 m_{(2)}^2 \left(\bm{p}_{(1)}\cdot\bm{n}\right){}^2}{4 m_{(1)}^2 r^6}
\right. \right. \right.
\nonumber\\
%
&\quad + \left.\left.\left.
\frac{9 m_{(2)} (\bm{p}_{(1)}\cdot\bm{p}_{(2)})}{2 m_{(1)} r^6}-\frac{15 m_{(2)}^2 \bm{p}_{(1)}^2}{8 m_{(1)}^2 r^6}+\frac{9 \left(\bm{p}_{(2)}\cdot\bm{n}\right){}^2}{2 r^6}-\frac{9 \bm{p}_{(2)}^2}{4 r^6}\right) \right] \right\}
- \left( 1\leftrightarrow 2 \right) \,,
\end{align}
and
\begin{align}
\bm{Y}^i =
\frac{1}{c^4}
\left[ -\frac{1}{4} G (\bm{p}_{(2)}\cdot\bm{n}) \bm{p}_{(1)}^i \right] + \left( 1\leftrightarrow 2 \right) \,.
\end{align}
The existence of this unique vector $\bm{G}$ provides us with a stringent consistency check on the adiabatic tidal Hamiltonian we derived.

\subsection{Binding energy for circular binaries}
In this section, we compute the binding energy in the COM frame for circular orbits. The gauge invariant relation between the binding energy and the orbital frequency for circular orbits ($p_r=0$) is obtained by eliminating the dependence on the radial coordinate.
For circular orbits we have
    $\partial \widetilde{\mathcal{H}}(\widetilde{r},\widetilde{L})/ \partial \widetilde{r}=0\, .
    $
We then proceed as follows. 
First, we invert this relation to express $\widetilde{r}$ as a function of $\widetilde{L}$. 
Next, we substitute $\widetilde{L}$, written as a function of the orbital frequency $\widetilde{\omega} = \partial \widetilde{\mathcal{H}}(\widetilde{L})/\partial \widetilde{L}$, in the Hamiltonian~\eqref{eq_ham_ad}.
Following this procedure we obtain the binding energy $E$ as,
\begin{align}\label{eq_BE}
E(x,\widetilde{\lambda}_{(a)})= E_{\text{pp}}(x)+E_{\text{AT}}(x,\widetilde{\lambda}_{(a)})\, ,
\end{align} 
where we introduced $x=\widetilde{\omega}^{2/3}$, $E_{\text{pp}}$ can be found in Ref.~\cite{Mandal:2022nty}, and 
\begin{align}
\label{eq_energy}
E_{\text{AT}}(x,\widetilde{\lambda}_{(a)}) &=  x^6 \left(\frac92\widetilde{\lambda}_{(+)}\right) + x^7 \left[  \left(-\frac{33  }{4}\nu+\frac{121}{8}\right)\widetilde{\lambda}_{(+)}+ \left( \frac{55}{8}\right) \delta \widetilde{\lambda}_{(-)} \right] \nonumber\\
&\quad + x^8 \left[ \left(\frac{91 }{16}\nu ^2-\frac{2717  }{42}\nu+\frac{20865}{224}\right) \widetilde{\lambda}_{(+)} + \left(-\frac{715  }{48}\nu+\frac{11583}{224}\right) \delta \widetilde{\lambda}_{(-)}  \right] \,,
\end{align}
where 
\begin{align}
\widetilde{\lambda}_{(\pm)} =
\frac{m_{(2)}}{m_{(1)}} \widetilde{\lambda}_{(1)}
\,\pm\,
\frac{m_{(1)}}{m_{(2)}} \widetilde{\lambda}_{(2)} \,,
\end{align}
This expression for the binding energy agrees with the previously known result of Ref.~\cite{Henry:2019xhg}, Eq.~(6.5b), derived using classical PN techniques~\cite{Fokker:1929,Damour:1985mt}.

\subsection{Scattering angle for hyperbolic encounters}

As a second application, we now compute the scattering angle $\chi$ in the COM frame for the hyperbolic encounter of two stars. 
To do this calculation we as follows. 
First, we re-express the Hamiltonian $\mathcal{H}$ (which is a function of $p_r$, $L$ and $r$) to obtain $p_r = p_r(\mathcal{H},L,r)$.
Next, we use relation between the Lorentz factor $\gamma$ and the total energy per  total rest mass $\Gamma = \mathcal{H}/(Mc^2)$ given by 
\begin{align}\label{eq_gamma_for_v}
\gamma=\frac{1}{\sqrt{1-v^2/c^2}}=1+\frac{\Gamma^2-1}{2\nu}\, ,
\end{align}
where $v\equiv|\dot{\bm{r}}|$ is the relative velocity of the compact objects, 
and the total angular momentum $L$ and the impact parameter $b$ are related by
$L=(\mu\gamma v b)/\Gamma$.
This allows us to exchange $\mathcal{H}$ for $v$ and $L$ for $b$. 
Put together, we can then write the scattering angle as
\begin{align}
\chi(v,b) = - \frac{\gamma}{\mu\gamma v} \, \int \dd r \, \frac{\partial p_r(v,b,r)}{\partial b}  - \pi\, .
\end{align}

Performing this procedure with the Hamiltonian~\eqref{eq_ham_ad} yields the scattering angle computed in the COM frame, which we write as
\begin{align}\label{eq_SA}
\chi(v,b)= \chi_{\text{pp}}(v,b)+\chi_{\text{AT}}(v,b) \,,
\end{align}
with the following adiabatic tidal contribution
\begin{align}
\label{eq:SA_ES2_N3LO}
\frac{\chi_{\rm AT}}{\Gamma} &=  \frac{v^2}{M b^4}
\begin{bmatrix}%
\lambda_{\rm (+)} & ~\delta \lambda_{\rm (-)}
\end{bmatrix}
\cdot \left\{  
\pi \left(\frac{G_NM}{v^2b}\right)^2\begin{bmatrix} 1\\ 0\end{bmatrix} \bigg\{  \frac{45}{16}  + \frac{135}{32}  \left(\frac{v^2}{c^2}\right) + \frac{1575}{256}  \left(\frac{v^4}{c^4}\right) \bigg\} \right. \nonumber\\
&\quad\left. + \left(\frac{G_NM}{v^2b}\right)^3 \left\{48 \begin{bmatrix} 1\\ 0\end{bmatrix}  + \begin{bmatrix} 732/5\\12 \end{bmatrix}  \left(\frac{v^2}{c^2}\right) + \frac{3}{35}\begin{bmatrix} 3073\\ 593\end{bmatrix} \left(\frac{v^4}{c^4}\right) \right\}
\right. \nonumber\\
&\quad\left. +\, \pi \, \left(\frac{G_NM}{v^2b}\right)^4 \left\{\frac{315}{8}\begin{bmatrix} 1\\ 0\end{bmatrix} +\frac{315}{64}\begin{bmatrix} 51-2\nu\\ 5\end{bmatrix}  \left(\frac{v^2}{c^2}\right) + \frac{15}{128}\begin{bmatrix} 5331-274\nu \\ 1383\end{bmatrix}  \left(\frac{v^4}{c^4}\right) \right\} \right\} 
\nonumber\\
&\quad +\mathcal{O}\left(G_N^5,\frac{v^6}{c^6}\right) \,,
\end{align}
where we introduced
\begin{align}
\lambda_{(\pm)}=\frac{m_{(2)}}{m_{(1)}} \, \lambda_{(1)} \,\pm\, \frac{m_{(1)}}{m_{(2)}} \, \lambda_{(2)} \,,
\end{align}
and $\chi_{\text{pp}}$ is reported in Ref.~\cite{Mandal:2022nty}, Section 6.2.
Notice that we use a matrix notation in Eq.~\eqref{eq:SA_ES2_N3LO} to make the expression shorter.
Equation~\eqref{eq:SA_ES2_N3LO} agrees to 3PM (i.e.,~$G_N^3$) with the result reported by Ref.~\cite{Jakobsen:2022psy,Kalin:2020lmz}, obtained using techniques of worldline QFT~\cite{Mogull:2020sak} and the EFT developed for PM calculations~\cite{Kalin:2020mvi}, respectively.
\section{Conclusions}\label{sec_Conclusion}

In this work, we derived an effective Hamiltonian that describes the dynamical gravitoelectric tidal interaction between two nonspinning compact objects up to the 2PN order. 
We also computed the effective Hamiltonian in the adiabatic limit, which we used 
to calculate two gauge-invariant quantities, namely, the binding energy of a circular binary and the scattering angle for a hyperbolic scattering.
These result extend previous results in the literature and agree in their 
particular limits.
We expect our result to be used to improve accuracy of gravitational waveform models,
for applications to present and future ground-based GW observatories.

Our work can be extended in several directions.
One possibility is to compute the next higher-order corrections to the dynamic and adiabatic Hamiltonians. This would be important to further improve GW models for NS binaries.
Here, we focused on the dynamical gravitoelectric quadrupolar tides, but our framework is general enough to be extended to obtain higher-order corrections for the dynamical gravitomagnetic tides~\cite{Gupta:2020lnv,Gupta:2023oyy}, as well as to incorporate higher order multipolar tides.
It could also be interesting to work at lower PN order, but including additional physics to the model. 
For example, the coupling of the oscillation modes of the NS with other degrees of freedom, such as its spin~\cite{Ho:1998hq,Steinhoff:2021dsn,Ma:2020rak}, or other oscillation modes~\cite{2012ApJ...751..136W,2013ApJ...769..121W,Landry:2015snx,Xu:2017hqo,Essick:2016tkn,Yu:2022fzw} could also be incorporated to improve the model's level of physical realism.
Finally, we could follow Refs.~\cite{Hinderer:2016eia,Steinhoff:2016rfi} and add our 2PN Hamiltonians into time-domain effective-one-body waveform models
to improve their agreement with numerical relativity simulations.

\subsection*{Acknowledgements}
We thank Quentin Henry, Gustav Jakobsen, Jung-Wook Kim, Saketh MVS, and Sebastian V\"{o}lkel for insightful discussions. We are particularly grateful to Quentin Henry who provided us the Hamiltonian obtained in Ref.~\cite{Henry:2019xhg}, which we used to validate
some of our results. We thank Marcus Haberland for pointing out a typo in equation \eqref{eq_energy}.
The figures in this work were produced with {\sc TikZ}~\cite{tantau:2021}.
The work of M.K.M is supported by Fellini~-~Fellowship for Innovation at INFN funded by the European Union's Horizon 2020 research and innovation programme under the Marie Sk{\l}odowska-Curie grant agreement No.~754496. 
H.O.S acknowledges funding from the Deutsche Forschungsgemeinschaft (DFG)~-~project number:~386119226.
R.P.'s research is funded by the Deutsche Forschungsgemeinschaft (DFG, German Research Foundation), Projektnummer 417533893/GRK2575 “Rethinking Quantum Field Theory”.
%

\bibliographystyle{JHEP}
\bibliography{biblio}

\providecommand{\href}[2]{#2}\begingroup\raggedright\begin{thebibliography}{100}

\bibitem{LIGOScientific:2017vwq}
{\bf LIGO Scientific, Virgo} Collaboration, B.~P. Abbott et~al., {\it
  {GW170817: Observation of Gravitational Waves from a Binary Neutron Star
  Inspiral}},  {\em Phys. Rev. Lett.} {\bf 119} (2017), no.~16 161101,
  [\href{http://arxiv.org/abs/1710.05832}{{\tt arXiv:1710.05832}}].

\bibitem{LIGOScientific:2017ync}
{\bf LIGO Scientific, Virgo, Fermi GBM, INTEGRAL, IceCube, AstroSat Cadmium
  Zinc Telluride Imager Team, IPN, Insight-Hxmt, ANTARES, Swift, AGILE Team,
  1M2H Team, Dark Energy Camera GW-EM, DES, DLT40, GRAWITA, Fermi-LAT, ATCA,
  ASKAP, Las Cumbres Observatory Group, OzGrav, DWF (Deeper Wider Faster
  Program), AST3, CAASTRO, VINROUGE, MASTER, J-GEM, GROWTH, JAGWAR,
  CaltechNRAO, TTU-NRAO, NuSTAR, Pan-STARRS, MAXI Team, TZAC Consortium, KU,
  Nordic Optical Telescope, ePESSTO, GROND, Texas Tech University, SALT Group,
  TOROS, BOOTES, MWA, CALET, IKI-GW Follow-up, H.E.S.S., LOFAR, LWA, HAWC,
  Pierre Auger, ALMA, Euro VLBI Team, Pi of Sky, Chandra Team at McGill
  University, DFN, ATLAS Telescopes, High Time Resolution Universe Survey,
  RIMAS, RATIR, SKA South Africa/MeerKAT} Collaboration, B.~P. Abbott et~al.,
  {\it {Multi-messenger Observations of a Binary Neutron Star Merger}},  {\em
  Astrophys. J. Lett.} {\bf 848} (2017), no.~2 L12,
  [\href{http://arxiv.org/abs/1710.05833}{{\tt arXiv:1710.05833}}].

\bibitem{LIGOScientific:2020aai}
{\bf LIGO Scientific, Virgo} Collaboration, B.~P. Abbott et~al., {\it
  {GW190425: Observation of a Compact Binary Coalescence with Total Mass $\sim
  3.4 M_{\odot}$}},  {\em Astrophys. J. Lett.} {\bf 892} (2020), no.~1 L3,
  [\href{http://arxiv.org/abs/2001.01761}{{\tt arXiv:2001.01761}}].

\bibitem{Flanagan:2007ix}
E.~E. Flanagan and T.~Hinderer, {\it {Constraining neutron star tidal Love
  numbers with gravitational wave detectors}},  {\em Phys. Rev. D} {\bf 77}
  (2008) 021502, [\href{http://arxiv.org/abs/0709.1915}{{\tt
  arXiv:0709.1915}}].

\bibitem{LIGOScientific:2018hze}
{\bf LIGO Scientific, Virgo} Collaboration, B.~P. Abbott et~al., {\it
  {Properties of the binary neutron star merger GW170817}},  {\em Phys. Rev. X}
  {\bf 9} (2019), no.~1 011001, [\href{http://arxiv.org/abs/1805.11579}{{\tt
  arXiv:1805.11579}}].

\bibitem{LIGOScientific:2018cki}
{\bf LIGO Scientific, Virgo} Collaboration, B.~P. Abbott et~al., {\it
  {GW170817: Measurements of neutron star radii and equation of state}},  {\em
  Phys. Rev. Lett.} {\bf 121} (2018), no.~16 161101,
  [\href{http://arxiv.org/abs/1805.11581}{{\tt arXiv:1805.11581}}].

\bibitem{Chatziioannou:2020pqz}
K.~Chatziioannou, {\it {Neutron star tidal deformability and equation of state
  constraints}},  {\em Gen. Rel. Grav.} {\bf 52} (2020), no.~11 109,
  [\href{http://arxiv.org/abs/2006.03168}{{\tt arXiv:2006.03168}}].

\bibitem{Pradhan:2022rxs}
B.~K. Pradhan, A.~Vijaykumar, and D.~Chatterjee, {\it {Impact of updated
  multipole Love numbers and f-Love universal relations in the context of
  binary neutron stars}},  {\em Phys. Rev. D} {\bf 107} (2023), no.~2 023010,
  [\href{http://arxiv.org/abs/2210.09425}{{\tt arXiv:2210.09425}}].

\bibitem{McDermott:1983ApJ...268..837M}
P.~N. {McDermott}, H.~M. {van Horn}, and J.~F. {Scholl}, {\it {Nonradial g-mode
  oscillations of warm neutron stars}},  {\em Astrophys. J.} {\bf 268} (1983)
  837--848.

\bibitem{McDermott:1988ApJ...325..725M}
P.~N. {McDermott}, H.~M. {van Horn}, and C.~J. {Hansen}, {\it {Nonradial
  Oscillations of Neutron Stars}},  {\em Astrophys. J.} {\bf 325} (1988) 725.

\bibitem{christensen1998lecture}
J.~Christensen-Dalsgaard, {\em Lecture Notes on Stellar Oscillations}.
\newblock Institut for Fysik og Astronomi, Aarhus Universitet, 1998.

\bibitem{Kokkotas:1999bd}
K.~D. Kokkotas and B.~G. Schmidt, {\it {Quasinormal modes of stars and black
  holes}},  {\em Living Rev. Rel.} {\bf 2} (1999) 2,
  [\href{http://arxiv.org/abs/gr-qc/9909058}{{\tt gr-qc/9909058}}].

\bibitem{Andersson:2021qdq}
N.~Andersson, {\it {A gravitational-wave perspective on neutron-star
  seismology}},  {\em Universe} {\bf 7} (2021), no.~4 97,
  [\href{http://arxiv.org/abs/2103.10223}{{\tt arXiv:2103.10223}}].

\bibitem{Osaki1975}
Y.~Osaki, {\it {Nonradial oscillations of a 10 solar mass star in the
  main-sequence stage}},  {\em Publications of the Astronomical Society of
  Japan} {\bf 27} (1975), no.~2 237--258.

\bibitem{Andersson:1995ez}
N.~Andersson, Y.~Kojima, and K.~D. Kokkotas, {\it {On the oscillation spectra
  of ultracompact stars: An Extensive survey of gravitational wave modes}},
  {\em Astrophys. J.} {\bf 462} (1996) 855,
  [\href{http://arxiv.org/abs/gr-qc/9512048}{{\tt gr-qc/9512048}}].

\bibitem{Andersson:1997rn}
N.~Andersson and K.~D. Kokkotas, {\it {Towards gravitational wave
  asteroseismology}},  {\em Mon. Not. Roy. Astron. Soc.} {\bf 299} (1998)
  1059--1068, [\href{http://arxiv.org/abs/gr-qc/9711088}{{\tt gr-qc/9711088}}].

\bibitem{Will:1983vlw}
C.~M. Will, {\it {Tidal gravitational radiation from homogeneous stars}},  {\em
  Astrophys. J.} {\bf 274} (1983) 858--874.

\bibitem{cowlingfmodes}
T.~G. Cowling, {\it {The Non-radial Oscillations of Polytropic Stars}},  {\em
  Monthly Notices of the Royal Astronomical Society} {\bf 101} (12, 1941)
  367--375.

\bibitem{shibatafmodes}
M.~Shibata, {\it {Effects of tidal resonances in coalescing compact binary
  systems}},  {\em Prog. Theor. Phys.} {\bf 91} (1994) 871--884.

\bibitem{Kokkotas:1995xe}
K.~D. Kokkotas and G.~Schaefer, {\it {Tidal and tidal resonant effects in
  coalescing binaries}},  {\em Mon. Not. Roy. Astron. Soc.} {\bf 275} (1995)
  301, [\href{http://arxiv.org/abs/gr-qc/9502034}{{\tt gr-qc/9502034}}].

\bibitem{Lai:1993di}
D.~Lai, {\it {Resonant oscillations and tidal heating in coalescing binary
  neutron stars}},  {\em Mon. Not. Roy. Astron. Soc.} {\bf 270} (1994) 611,
  [\href{http://arxiv.org/abs/astro-ph/9404062}{{\tt astro-ph/9404062}}].

\bibitem{Ho:1998hq}
W.~C.~G. Ho and D.~Lai, {\it {Resonant tidal excitations of rotating neutron
  stars in coalescing binaries}},  {\em Mon. Not. Roy. Astron. Soc.} {\bf 308}
  (1999) 153, [\href{http://arxiv.org/abs/astro-ph/9812116}{{\tt
  astro-ph/9812116}}].

\bibitem{Lai:2006pr}
D.~Lai and Y.~Wu, {\it {Resonant Tidal Excitations of Inertial Modes in
  Coalescing Neutron Star Binaries}},  {\em Phys. Rev. D} {\bf 74} (2006)
  024007, [\href{http://arxiv.org/abs/astro-ph/0604163}{{\tt
  astro-ph/0604163}}].

\bibitem{poisson2014gravity}
E.~Poisson and C.~Will, {\em Gravity: Newtonian, Post-Newtonian, Relativistic}.
\newblock Cambridge University Press, 2014.

\bibitem{Hinderer:2007mb}
T.~Hinderer, {\it {Tidal Love numbers of neutron stars}},  {\em Astrophys. J.}
  {\bf 677} (2008) 1216--1220, [\href{http://arxiv.org/abs/0711.2420}{{\tt
  arXiv:0711.2420}}].

\bibitem{Binnington:2009bb}
T.~Binnington and E.~Poisson, {\it {Relativistic theory of tidal Love
  numbers}},  {\em Phys. Rev. D} {\bf 80} (2009) 084018,
  [\href{http://arxiv.org/abs/0906.1366}{{\tt arXiv:0906.1366}}].

\bibitem{Damour:2009vw}
T.~Damour and A.~Nagar, {\it {Relativistic tidal properties of neutron stars}},
   {\em Phys. Rev. D} {\bf 80} (2009) 084035,
  [\href{http://arxiv.org/abs/0906.0096}{{\tt arXiv:0906.0096}}].

\bibitem{Love:1911}
A.~E.~H. Love, {\it {Some problems of geodynamics}},  {\em Cambridge U. Press}
  (1911).

\bibitem{Yagi:2016ejg}
K.~Yagi and N.~Yunes, {\it {I-Love-Q Relations: From Compact Stars to Black
  Holes}},  {\em Class. Quant. Grav.} {\bf 33} (2016), no.~9 095005,
  [\href{http://arxiv.org/abs/1601.02171}{{\tt arXiv:1601.02171}}].

\bibitem{Steinhoff:2016rfi}
J.~Steinhoff, T.~Hinderer, A.~Buonanno, and A.~Taracchini, {\it {Dynamical
  Tides in General Relativity: Effective Action and Effective-One-Body
  Hamiltonian}},  {\em Phys. Rev. D} {\bf 94} (2016), no.~10 104028,
  [\href{http://arxiv.org/abs/1608.01907}{{\tt arXiv:1608.01907}}].

\bibitem{Bini:2012gu}
D.~Bini, T.~Damour, and G.~Faye, {\it {Effective action approach to
  higher-order relativistic tidal interactions in binary systems and their
  effective one body description}},  {\em Phys. Rev. D} {\bf 85} (2012) 124034,
  [\href{http://arxiv.org/abs/1202.3565}{{\tt arXiv:1202.3565}}].

\bibitem{Thorne:1984mz}
K.~S. Thorne and J.~B. Hartle, {\it {Laws of motion and precession for black
  holes and other bodies}},  {\em Phys. Rev. D} {\bf 31} (1984) 1815--1837.

\bibitem{Zhang:1986cpa}
X.~H. Zhang, {\it {Multipole expansions of the general-relativistic
  gravitational field of the external universe}},  {\em Phys. Rev. D} {\bf 34}
  (1986), no.~4 991--1004.

\bibitem{Damour:1990pi}
T.~Damour, M.~Soffel, and C.-m. Xu, {\it {General relativistic celestial
  mechanics. 1. Method and definition of reference systems}},  {\em Phys. Rev.
  D} {\bf 43} (1991) 3273--3307.

\bibitem{Damour:2009wj}
T.~Damour and A.~Nagar, {\it {Effective One Body description of tidal effects
  in inspiralling compact binaries}},  {\em Phys. Rev. D} {\bf 81} (2010)
  084016, [\href{http://arxiv.org/abs/0911.5041}{{\tt arXiv:0911.5041}}].

\bibitem{Henry:2019xhg}
Q.~Henry, G.~Faye, and L.~Blanchet, {\it {Tidal effects in the equations of
  motion of compact binary systems to next-to-next-to-leading post-Newtonian
  order}},  {\em Phys. Rev. D} {\bf 101} (2020), no.~6 064047,
  [\href{http://arxiv.org/abs/1912.01920}{{\tt arXiv:1912.01920}}].

\bibitem{Favata:2005da}
M.~Favata, {\it {Are neutron stars crushed? gravitomagnetic tidal fields as a
  mechanism for binary-induced collapse}},  {\em Phys. Rev. D} {\bf 73} (2006)
  104005, [\href{http://arxiv.org/abs/astro-ph/0510668}{{\tt
  astro-ph/0510668}}].

\bibitem{Landry:2015cva}
P.~Landry and E.~Poisson, {\it {Gravitomagnetic response of an irrotational
  body to an applied tidal field}},  {\em Phys. Rev. D} {\bf 91} (2015), no.~10
  104026, [\href{http://arxiv.org/abs/1504.06606}{{\tt arXiv:1504.06606}}].

\bibitem{Pani:2018inf}
P.~Pani, L.~Gualtieri, T.~Abdelsalhin, and X.~Jim\'enez-Forteza, {\it {Magnetic
  tidal Love numbers clarified}},  {\em Phys. Rev. D} {\bf 98} (2018), no.~12
  124023, [\href{http://arxiv.org/abs/1810.01094}{{\tt arXiv:1810.01094}}].

\bibitem{Banihashemi:2018xfb}
B.~Banihashemi and J.~Vines, {\it {Gravitomagnetic tidal effects in
  gravitational waves from neutron star binaries}},  {\em Phys. Rev. D} {\bf
  101} (2020), no.~6 064003, [\href{http://arxiv.org/abs/1805.07266}{{\tt
  arXiv:1805.07266}}].

\bibitem{Poisson:2020eki}
E.~Poisson, {\it {Gravitomagnetic tidal resonance in neutron-star binary
  inspirals}},  {\em Phys. Rev. D} {\bf 101} (2020), no.~10 104028,
  [\href{http://arxiv.org/abs/2003.10427}{{\tt arXiv:2003.10427}}].

\bibitem{Poisson:2020mdi}
E.~Poisson, {\it {Gravitomagnetic Love tensor of a slowly rotating body:
  post-Newtonian theory}},  {\em Phys. Rev. D} {\bf 102} (2020), no.~6 064059,
  [\href{http://arxiv.org/abs/2007.01678}{{\tt arXiv:2007.01678}}].

\bibitem{Poisson:2020ify}
E.~Poisson and C.~Buisson, {\it {Tidal driving of inertial modes of Maclaurin
  spheroids}},  {\em Phys. Rev. D} {\bf 102} (2020), no.~10 104005,
  [\href{http://arxiv.org/abs/2007.03050}{{\tt arXiv:2007.03050}}].

\bibitem{Gupta:2020lnv}
P.~K. Gupta, J.~Steinhoff, and T.~Hinderer, {\it {Relativistic effective action
  of dynamical gravitomagnetic tides for slowly rotating neutron stars}},  {\em
  Phys. Rev. Res.} {\bf 3} (2021), no.~1 013147,
  [\href{http://arxiv.org/abs/2011.03508}{{\tt arXiv:2011.03508}}].

\bibitem{Pratten:2021pro}
G.~Pratten, P.~Schmidt, and N.~Williams, {\it {Impact of Dynamical Tides on the
  Reconstruction of the Neutron Star Equation of State}},  {\em Phys. Rev.
  Lett.} {\bf 129} (2022), no.~8 081102,
  [\href{http://arxiv.org/abs/2109.07566}{{\tt arXiv:2109.07566}}].

\bibitem{Hinderer:2016eia}
T.~Hinderer et~al., {\it {Effects of neutron-star dynamic tides on
  gravitational waveforms within the effective-one-body approach}},  {\em Phys.
  Rev. Lett.} {\bf 116} (2016), no.~18 181101,
  [\href{http://arxiv.org/abs/1602.00599}{{\tt arXiv:1602.00599}}].

\bibitem{Andersson:2019dwg}
N.~Andersson and P.~Pnigouras, {\it {The phenomenology of dynamical neutron
  star tides}},  {\em Mon. Not. Roy. Astron. Soc.} {\bf 503} (2021), no.~1
  533--539, [\href{http://arxiv.org/abs/1905.00012}{{\tt arXiv:1905.00012}}].

\bibitem{Schmidt:2019wrl}
P.~Schmidt and T.~Hinderer, {\it {Frequency domain model of $f$-mode dynamic
  tides in gravitational waveforms from compact binary inspirals}},  {\em Phys.
  Rev. D} {\bf 100} (2019), no.~2 021501,
  [\href{http://arxiv.org/abs/1905.00818}{{\tt arXiv:1905.00818}}].

\bibitem{Punturo:2010zza}
M.~Punturo et~al., {\it {The third generation of gravitational wave
  observatories and their science reach}},  {\em Class. Quant. Grav.} {\bf 27}
  (2010) 084007.

\bibitem{Reitze:2019iox}
D.~Reitze et~al., {\it {Cosmic Explorer: The U.S. Contribution to
  Gravitational-Wave Astronomy beyond LIGO}},  {\em Bull. Am. Astron. Soc.}
  {\bf 51} (2019), no.~7 035, [\href{http://arxiv.org/abs/1907.04833}{{\tt
  arXiv:1907.04833}}].

\bibitem{Ackley:2020atn}
K.~Ackley et~al., {\it {Neutron Star Extreme Matter Observatory: A
  kilohertz-band gravitational-wave detector in the global network}},  {\em
  Publ. Astron. Soc. Austral.} {\bf 37} (2020) e047,
  [\href{http://arxiv.org/abs/2007.03128}{{\tt arXiv:2007.03128}}].

\bibitem{Goldberger:2004jt}
W.~D. Goldberger and I.~Z. Rothstein, {\it {An Effective field theory of
  gravity for extended objects}},  {\em Phys. Rev. D} {\bf 73} (2006) 104029,
  [\href{http://arxiv.org/abs/hep-th/0409156}{{\tt hep-th/0409156}}].

\bibitem{Blanchet:2013haa}
L.~Blanchet, {\it {Gravitational Radiation from Post-Newtonian Sources and
  Inspiralling Compact Binaries}},  {\em Living Rev. Rel.} {\bf 17} (2014) 2,
  [\href{http://arxiv.org/abs/1310.1528}{{\tt arXiv:1310.1528}}].

\bibitem{Kol:2013ega}
B.~Kol and R.~Shir, {\it {Classical 3-loop 2-body diagrams}},  {\em JHEP} {\bf
  09} (2013) 069, [\href{http://arxiv.org/abs/1306.3220}{{\tt
  arXiv:1306.3220}}].

\bibitem{Foffa:2016rgu}
S.~Foffa, P.~Mastrolia, R.~Sturani, and C.~Sturm, {\it {Effective field theory
  approach to the gravitational two-body dynamics, at fourth post-Newtonian
  order and quintic in the Newton constant}},  {\em Phys. Rev. D} {\bf 95}
  (2017), no.~10 104009, [\href{http://arxiv.org/abs/1612.00482}{{\tt
  arXiv:1612.00482}}].

\bibitem{Mandal:2022nty}
M.~K. Mandal, P.~Mastrolia, R.~Patil, and J.~Steinhoff, {\it {Gravitational
  Spin-Orbit Hamiltonian at NNNLO in the post-Newtonian framework}},
  \href{http://arxiv.org/abs/2209.00611}{{\tt arXiv:2209.00611}}.

\bibitem{Mandal:2022ufb}
M.~K. Mandal, P.~Mastrolia, R.~Patil, and J.~Steinhoff, {\it {Gravitational
  Quadratic-in-Spin Hamiltonian at NNNLO in the post-Newtonian framework}},
  \href{http://arxiv.org/abs/2210.09176}{{\tt arXiv:2210.09176}}.

\bibitem{Porto:2016pyg}
R.~A. Porto, {\it {The effective field theorist\textquoteright{}s approach to
  gravitational dynamics}},  {\em Phys. Rept.} {\bf 633} (2016) 1--104,
  [\href{http://arxiv.org/abs/1601.04914}{{\tt arXiv:1601.04914}}].

\bibitem{Levi:2018nxp}
M.~Levi, {\it {Effective Field Theories of Post-Newtonian Gravity: A
  comprehensive review}},  {\em Rept. Prog. Phys.} {\bf 83} (2020), no.~7
  075901, [\href{http://arxiv.org/abs/1807.01699}{{\tt arXiv:1807.01699}}].

\bibitem{Goldberger:2022ebt}
W.~D. Goldberger, {\it {Effective field theories of gravity and compact binary
  dynamics: A Snowmass 2021 whitepaper}},  in {\em {Snowmass 2021}}, 6, 2022.
\newblock \href{http://arxiv.org/abs/2206.14249}{{\tt arXiv:2206.14249}}.

\bibitem{Vines:2010ca}
J.~E. Vines and E.~E. Flanagan, {\it {Post-1-Newtonian quadrupole tidal
  interactions in binary systems}},  {\em Phys. Rev. D} {\bf 88} (2013) 024046,
  [\href{http://arxiv.org/abs/1009.4919}{{\tt arXiv:1009.4919}}].

\bibitem{Steinhoff:2021dsn}
J.~Steinhoff, T.~Hinderer, T.~Dietrich, and F.~Foucart, {\it {Spin effects on
  neutron star fundamental-mode dynamical tides: Phenomenology and comparison
  to numerical simulations}},  {\em Phys. Rev. Res.} {\bf 3} (2021), no.~3
  033129, [\href{http://arxiv.org/abs/2103.06100}{{\tt arXiv:2103.06100}}].

\bibitem{Gupta:2023oyy}
P.~K. Gupta, J.~Steinhoff, and T.~Hinderer, {\it {Effect of dynamical
  gravitomagnetic tides on measurability of tidal parameters for binary neutron
  stars using gravitational waves}},
  \href{http://arxiv.org/abs/2302.11274}{{\tt arXiv:2302.11274}}.

\bibitem{Vines:2011ud}
J.~Vines, E.~E. Flanagan, and T.~Hinderer, {\it {Post-1-Newtonian tidal effects
  in the gravitational waveform from binary inspirals}},  {\em Phys. Rev. D}
  {\bf 83} (2011) 084051, [\href{http://arxiv.org/abs/1101.1673}{{\tt
  arXiv:1101.1673}}].

\bibitem{Abdelsalhin:2018reg}
T.~Abdelsalhin, L.~Gualtieri, and P.~Pani, {\it {Post-Newtonian spin-tidal
  couplings for compact binaries}},  {\em Phys. Rev. D} {\bf 98} (2018) 104046,
  [\href{http://arxiv.org/abs/1805.01487}{{\tt arXiv:1805.01487}}].

\bibitem{Landry:2018bil}
P.~Landry, {\it {Rotational-tidal phasing of the binary neutron star
  waveform}},  \href{http://arxiv.org/abs/1805.01882}{{\tt arXiv:1805.01882}}.

\bibitem{Jakobsen:2022psy}
G.~U. Jakobsen, G.~Mogull, J.~Plefka, and B.~Sauer, {\it {All things retarded:
  radiation-reaction in worldline quantum field theory}},  {\em JHEP} {\bf 10}
  (2022) 128, [\href{http://arxiv.org/abs/2207.00569}{{\tt arXiv:2207.00569}}].

\bibitem{Kalin:2020lmz}
G.~K\"alin, Z.~Liu, and R.~A. Porto, {\it {Conservative Tidal Effects in
  Compact Binary Systems to Next-to-Leading Post-Minkowskian Order}},  {\em
  Phys. Rev. D} {\bf 102} (2020) 124025,
  [\href{http://arxiv.org/abs/2008.06047}{{\tt arXiv:2008.06047}}].

\bibitem{Kalin:2020mvi}
G.~K\"alin and R.~A. Porto, {\it {Post-Minkowskian Effective Field Theory for
  Conservative Binary Dynamics}},  {\em JHEP} {\bf 11} (2020) 106,
  [\href{http://arxiv.org/abs/2006.01184}{{\tt arXiv:2006.01184}}].

\bibitem{Bini:2020flp}
D.~Bini, T.~Damour, and A.~Geralico, {\it {Scattering of tidally interacting
  bodies in post-Minkowskian gravity}},  {\em Phys. Rev. D} {\bf 101} (2020),
  no.~4 044039, [\href{http://arxiv.org/abs/2001.00352}{{\tt
  arXiv:2001.00352}}].

\bibitem{Cheung:2020sdj}
C.~Cheung and M.~P. Solon, {\it {Tidal Effects in the Post-Minkowskian
  Expansion}},  {\em Phys. Rev. Lett.} {\bf 125} (2020), no.~19 191601,
  [\href{http://arxiv.org/abs/2006.06665}{{\tt arXiv:2006.06665}}].

\bibitem{Haddad:2020que}
K.~Haddad and A.~Helset, {\it {Tidal effects in quantum field theory}},  {\em
  JHEP} {\bf 12} (2020) 024, [\href{http://arxiv.org/abs/2008.04920}{{\tt
  arXiv:2008.04920}}].

\bibitem{Cheung:2020gbf}
C.~Cheung, N.~Shah, and M.~P. Solon, {\it {Mining the Geodesic Equation for
  Scattering Data}},  {\em Phys. Rev. D} {\bf 103} (2021), no.~2 024030,
  [\href{http://arxiv.org/abs/2010.08568}{{\tt arXiv:2010.08568}}].

\bibitem{Bern:2020uwk}
Z.~Bern, J.~Parra-Martinez, R.~Roiban, E.~Sawyer, and C.-H. Shen, {\it {Leading
  Nonlinear Tidal Effects and Scattering Amplitudes}},  {\em JHEP} {\bf 05}
  (2021) 188, [\href{http://arxiv.org/abs/2010.08559}{{\tt arXiv:2010.08559}}].

\bibitem{Mougiakakos:2022sic}
S.~Mougiakakos, M.~M. Riva, and F.~Vernizzi, {\it {Gravitational Bremsstrahlung
  with Tidal Effects in the Post-Minkowskian Expansion}},  {\em Phys. Rev.
  Lett.} {\bf 129} (2022), no.~12 121101,
  [\href{http://arxiv.org/abs/2204.06556}{{\tt arXiv:2204.06556}}].

\bibitem{Buonanno:1998gg}
A.~Buonanno and T.~Damour, {\it {Effective one-body approach to general
  relativistic two-body dynamics}},  {\em Phys. Rev. D} {\bf 59} (1999) 084006,
  [\href{http://arxiv.org/abs/gr-qc/9811091}{{\tt gr-qc/9811091}}].

\bibitem{Buonanno:2000ef}
A.~Buonanno and T.~Damour, {\it {Transition from inspiral to plunge in binary
  black hole coalescences}},  {\em Phys. Rev. D} {\bf 62} (2000) 064015,
  [\href{http://arxiv.org/abs/gr-qc/0001013}{{\tt gr-qc/0001013}}].

\bibitem{Baiotti:2011am}
L.~Baiotti, T.~Damour, B.~Giacomazzo, A.~Nagar, and L.~Rezzolla, {\it {Accurate
  numerical simulations of inspiralling binary neutron stars and their
  comparison with effective-one-body analytical models}},  {\em Phys. Rev. D}
  {\bf 84} (2011) 024017, [\href{http://arxiv.org/abs/1103.3874}{{\tt
  arXiv:1103.3874}}].

\bibitem{Bernuzzi:2012ci}
S.~Bernuzzi, A.~Nagar, M.~Thierfelder, and B.~Brugmann, {\it {Tidal effects in
  binary neutron star coalescence}},  {\em Phys. Rev. D} {\bf 86} (2012)
  044030, [\href{http://arxiv.org/abs/1205.3403}{{\tt arXiv:1205.3403}}].

\bibitem{Bini:2014zxa}
D.~Bini and T.~Damour, {\it {Gravitational self-force corrections to two-body
  tidal interactions and the effective one-body formalism}},  {\em Phys. Rev.
  D} {\bf 90} (2014), no.~12 124037,
  [\href{http://arxiv.org/abs/1409.6933}{{\tt arXiv:1409.6933}}].

\bibitem{Collins:1984xc}
J.~C. Collins, {\em {Renormalization}: {An Introduction to Renormalization, The
  Renormalization Group, and the Operator Product Expansion}}, vol.~26 of {\em
  Cambridge Monographs on Mathematical Physics}.
\newblock Cambridge University Press, Cambridge, 1986.

\bibitem{Beneke:1997zp}
M.~Beneke and V.~A. Smirnov, {\it {Asymptotic expansion of Feynman integrals
  near threshold}},  {\em Nucl. Phys. B} {\bf 522} (1998) 321--344,
  [\href{http://arxiv.org/abs/hep-ph/9711391}{{\tt hep-ph/9711391}}].

\bibitem{Kol:2007bc}
B.~Kol and M.~Smolkin, {\it {Non-Relativistic Gravitation: From Newton to
  Einstein and Back}},  {\em Class. Quant. Grav.} {\bf 25} (2008) 145011,
  [\href{http://arxiv.org/abs/0712.4116}{{\tt arXiv:0712.4116}}].

\bibitem{Kol:2007rx}
B.~Kol and M.~Smolkin, {\it {Classical Effective Field Theory and Caged Black
  Holes}},  {\em Phys. Rev. D} {\bf 77} (2008) 064033,
  [\href{http://arxiv.org/abs/0712.2822}{{\tt arXiv:0712.2822}}].

\bibitem{Levi:2017kzq}
M.~Levi and J.~Steinhoff, {\it {EFTofPNG: A package for high precision
  computation with the Effective Field Theory of Post-Newtonian Gravity}},
  {\em Class. Quant. Grav.} {\bf 34} (2017), no.~24 244001,
  [\href{http://arxiv.org/abs/1705.06309}{{\tt arXiv:1705.06309}}].

\bibitem{xAct}
J.~M.~M. Garc\'ia, ``xact: Efficient tensor computer algebra for mathematica.''

\bibitem{Lee:2013mka}
R.~N. Lee, {\it {LiteRed 1.4: a powerful tool for reduction of multiloop
  integrals}},  {\em J. Phys. Conf. Ser.} {\bf 523} (2014) 012059,
  [\href{http://arxiv.org/abs/1310.1145}{{\tt arXiv:1310.1145}}].

\bibitem{Levi:2015msa}
M.~Levi and J.~Steinhoff, {\it {Spinning gravitating objects in the effective
  field theory in the post-Newtonian scheme}},  {\em JHEP} {\bf 09} (2015) 219,
  [\href{http://arxiv.org/abs/1501.04956}{{\tt arXiv:1501.04956}}].

\bibitem{Damour:2000kk}
T.~Damour, P.~Jaranowski, and G.~Schaefer, {\it {Poincare invariance in the ADM
  Hamiltonian approach to the general relativistic two-body problem}},  {\em
  Phys. Rev. D} {\bf 62} (2000) 021501,
  [\href{http://arxiv.org/abs/gr-qc/0003051}{{\tt gr-qc/0003051}}]. [Erratum:
  Phys.Rev.D 63, 029903 (2001)].

\bibitem{Levi:2016ofk}
M.~Levi and J.~Steinhoff, {\it {Complete conservative dynamics for inspiralling
  compact binaries with spins at the fourth post-Newtonian order}},  {\em JCAP}
  {\bf 09} (2021) 029, [\href{http://arxiv.org/abs/1607.04252}{{\tt
  arXiv:1607.04252}}].

\bibitem{Fokker:1929}
A.~D. Fokker, {\it {}},  {\em Z. Phys.} {\bf 58} (1929) 386.

\bibitem{Damour:1985mt}
T.~Damour and G.~Sch\"afer, {\it {Lagrangians for$n$ point masses at the second
  post-Newtonian approximation of general relativity}},  {\em Gen. Rel. Grav.}
  {\bf 17} (1985) 879--905.

\bibitem{Mogull:2020sak}
G.~Mogull, J.~Plefka, and J.~Steinhoff, {\it {Classical black hole scattering
  from a worldline quantum field theory}},  {\em JHEP} {\bf 02} (2021) 048,
  [\href{http://arxiv.org/abs/2010.02865}{{\tt arXiv:2010.02865}}].

\bibitem{Ma:2020rak}
S.~Ma, H.~Yu, and Y.~Chen, {\it {Excitation of f-modes during mergers of
  spinning binary neutron star}},  {\em Phys. Rev. D} {\bf 101} (2020), no.~12
  123020, [\href{http://arxiv.org/abs/2003.02373}{{\tt arXiv:2003.02373}}].

\bibitem{2012ApJ...751..136W}
N.~N. Weinberg, P.~Arras, E.~Quataert, and J.~Burkart, {\it {Nonlinear Tides in
  Close Binary Systems}},  {\em Astrophys. J.} {\bf 751} (2012) 136,
  [\href{http://arxiv.org/abs/1107.0946}{{\tt arXiv:1107.0946}}].

\bibitem{2013ApJ...769..121W}
N.~N. Weinberg, P.~Arras, and J.~Burkart, {\it {An instability due to the
  nonlinear coupling of p-modes to g-modes: Implications for coalescing neutron
  star binaries}},  {\em Astrophys. J.} {\bf 769} (2013) 121,
  [\href{http://arxiv.org/abs/1302.2292}{{\tt arXiv:1302.2292}}].

\bibitem{Landry:2015snx}
P.~Landry and E.~Poisson, {\it {Dynamical response to a stationary tidal
  field}},  {\em Phys. Rev. D} {\bf 92} (2015), no.~12 124041,
  [\href{http://arxiv.org/abs/1510.09170}{{\tt arXiv:1510.09170}}].

\bibitem{Xu:2017hqo}
W.~Xu and D.~Lai, {\it {Resonant Tidal Excitation of Oscillation Modes in
  Merging Binary Neutron Stars: Inertial-Gravity Modes}},  {\em Phys. Rev. D}
  {\bf 96} (2017), no.~8 083005, [\href{http://arxiv.org/abs/1708.01839}{{\tt
  arXiv:1708.01839}}].

\bibitem{Essick:2016tkn}
R.~Essick, S.~Vitale, and N.~N. Weinberg, {\it {Impact of the tidal p-g
  instability on the gravitational wave signal from coalescing binary neutron
  stars}},  {\em Phys. Rev. D} {\bf 94} (2016), no.~10 103012,
  [\href{http://arxiv.org/abs/1609.06362}{{\tt arXiv:1609.06362}}].

\bibitem{Yu:2022fzw}
H.~Yu, N.~N. Weinberg, P.~Arras, J.~Kwon, and T.~Venumadhav, {\it {Beyond the
  linear tide: impact of the non-linear tidal response of neutron stars on
  gravitational waveforms from binary inspirals}},  {\em Mon. Not. Roy. Astron.
  Soc.} {\bf 519} (2023), no.~3 4325--4343,
  [\href{http://arxiv.org/abs/2211.07002}{{\tt arXiv:2211.07002}}].

\bibitem{tantau:2021}
T.~Tantau, {\em The TikZ and PGF Packages}.

\end{thebibliography}\endgroup

\end{document}